\documentclass[12pt]{article}
\pdfoutput=1
\usepackage{epsf,amsfonts,amssymb,epsfig,amsmath,graphics,slashed}
\usepackage{hyperref,hep,graphicx,subfig}

\newcommand{\nn}{\notag \\}

\begin{document}

\makeatletter
\renewcommand{\theequation}{\thesection.\arabic{equation}}
\@addtoreset{equation}{section}
\makeatother

\baselineskip 18pt

\begin{titlepage}

\vfill

\begin{flushright}
Imperial/TP/2011/JG/07\\
\end{flushright}

\vfill

\begin{center}
   \baselineskip=16pt
   {\Large\bf Magnetic and electric $AdS$ solutions\\ in string- and M-theory}
  \vskip 1.5cm
      Aristomenis Donos, Jerome P. Gauntlett and Christiana Pantelidou\\
   \vskip .6cm
      \begin{small}
      \textit{Blackett Laboratory, 
        Imperial College\\ London, SW7 2AZ, U.K.}
        \end{small}\\*[.6cm]

\end{center}

\vfill

\begin{center}
\textbf{Abstract}
\end{center}

\begin{quote}
The stability properties of a family of magnetic $AdS_{3}\times \mathbb{R}^{2}$ solutions of $D=5$, $SO(6)$ gauged SUGRA
are investigated in more detail. We construct an analogous family of magnetic $AdS_{2}\times \mathbb{R}^{2}$ 
solutions of  $D=4$, $SO(8)$ gauged SUGRA, 
including a family of supersymmetric solutions, and also investigate their stability.
We construct supersymmetric domain walls that interpolate between $AdS_5$ and an $AdS_3\times\mathbb{R}^2$ solution
and also between $AdS_4$ and an $AdS_2\times\mathbb{R}^2$ solution which provide stable zero temperature
ground states for the corresponding dual CFTs. We also construct new families of electric $AdS_{2}\times \mathbb{R}^{3}$ and
$AdS_{2}\times \mathbb{R}^{2}$ solutions.
\end{quote}

\vfill

\end{titlepage}
\setcounter{equation}{0}


\section{Introduction}

With a view towards condensed matter applications, there have been several holographic investigations into
the behaviour of strongly coupled gauge theories in the presence of 
magnetic fields, starting with \cite{Hartnoll:2007ai,Hartnoll:2007ih,Hartnoll:2007ip,Albash:2008eh}.
The main focus of this paper will be on $AdS_3\times\mathbb{R}^{2}$ and $AdS_2\times\mathbb{R}^{2}$
solutions of string or M-theory that are supported by purely magnetic fields in the $\mathbb{R}^{2}$ directions. 
Such solutions are of interest because they provide
candidate holographic dual descriptions of the IR limit of the zero temperature ground states of field theories in $d=4$ and $d=3$, respectively. 
For this to be the case, it is certainly necessary 
that
the solutions are stable and, in particular, do not contain any modes that violate the $AdS_n$ BF bound. Such stability is
guaranteed if the solutions are supersymmetric.

As far as we are aware, the first constructions of such supersymmetric $AdS_3\times\mathbb{R}^2$ solutions
were presented in 
\cite{Gauntlett:2007sm}.
The solutions were obtained by uplifting a ``magnetovac" solution of Romans' $D=5$ gauged supergravity \cite{Romans:1985ps}
either on an $S^5$, to obtain a solution of type IIB supergravity, 
or on the general class of $M_6$ \cite{Lin:2004nb,Gaiotto:2009gz} 
associated with $AdS_5\times M_6$ solutions of $D=11$ supergravity dual to $N=2$ SCFTs in $d=4$ \cite{Gauntlett:2007sm}.

Subsequently, a non-supersymmetric magnetic $AdS_3\times\mathbb{R}^2$ solution of minimal gauged supergravity was shown to arise
as the near-horizon geometry of magnetic black brane solutions at zero temperature\footnote{Dyonic extensions were studied in \cite{D'Hoker:2009bc,D'Hoker:2010rz,D'Hoker:2010ij}.} \cite{D'Hoker:2009mm}.
These black brane solutions can again be uplifted on an $S^5$, to obtain a solution of type IIB supergravity, but also 
on general $X_5$ \cite{Gauntlett:2005ww} and $N_6$ \cite{Gauntlett:2004zh} associated with
$AdS_5\times X_5$ and $AdS_5\times N_6$ solutions of type IIB and $D=11$ supergravity, respectively, dual to $N=1$ SCFTs in $d=4$
\cite{Gauntlett:2006ai,Gauntlett:2007ma}.
One of the conclusions of this paper is that for the special case of uplifting on the $S^5$, for which the dual field theory is just $N=4$ SYM, 
the $AdS_3\times\mathbb{R}^2$ solution studied in \cite{D'Hoker:2009mm} is not stable and hence does not describe the zero temperature ground state.

More recently, it was found that these supersymmetric and non-supersymmetric magnetic $AdS_3\times\mathbb{R}^2$ solutions 
are 
members of a more general two-parameter family
of solutions \cite{Almuhairi:2010rb}, with the supersymmetric solution of \cite{Romans:1985ps,Gauntlett:2007sm} part of a one parameter sub-family, which can
be constructed within a $U(1)^3$
truncation of $D=5$ $SO(6)$ gauged supergravity.
An initial investigation into the stability of the non-supersymmetric $AdS_3\times\mathbb{R}^2$ solutions of \cite{Almuhairi:2010rb} was undertaken in 
\cite{Almuhairi:2011ws}, within the context of $SO(6)$ gauged supergravity.
Here we will re-examine the analysis of \cite{Almuhairi:2011ws} finding results which differ in some respects due to a mixing of modes that was
overlooked in \cite{Almuhairi:2011ws}.
In addition, we will show that a large class of the $AdS_3\times\mathbb{R}^2$ 
solutions also suffer from a new kind of instability involving neutral scalar fields that are spatially modulated in the $\mathbb{R}^2$ directions
similar to \cite{Nakamura:2009tf,Donos:2011bh,Bergman:2011rf,Donos:2011ff}.

Our results, and those of \cite{Almuhairi:2011ws}, imply that when the solutions are uplifted on $S^5$ to give type IIB solutions dual to $N=4$ SYM theory, the parameter space of potentially stable $AdS_3\times\mathbb{R}^2$ solutions is 
now very small, but still non-zero (see figures \ref{fig:susy} and \ref{fig:AdS3}). While we think it is unlikely that there are any further instabilities within $SO(6)$ gauged supergravity,
it is still possible that there others within the full KK spectrum. 
Note that if we consider the one-parameter family of solutions that lie within Romans' theory, some of
the instabilities that we discuss here, but not all, involve fields lying outside of Romans' theory.
This is relevant when we uplift the relevant solutions not an $S^5$ to $D=10$, but on the $M_6$ of \cite{Lin:2004nb} to $D=11$ \cite{Gauntlett:2007sm}.
Within minimal gauged supergravity, we do not find any instabilities for the unique magnetic $AdS_3\times\mathbb{R}^2$ solution.

We will also construct a supersymmetric domain wall solution that interpolates between $AdS_5$ in the UV and a particular
supersymmetric $AdS_3\times\mathbb{R}^2$
solution in the IR. The example we choose lies within Romans' theory so this can be uplifted both on $S^5$ to type IIB and also on the class of $M_6$ of
\cite{Lin:2004nb}  
to $D=11$. This solution, being supersymmetric, should describe the stable ultimate zero temperature ground state of the corresponding 
$d=4$ CFTs
when they are placed in a magnetic field.

The instabilities that we find for the non-supersymmetric $AdS_3\times\mathbb{R}^2$ solutions show that these solutions
cannot provide the ultimate IR ground states of dual $d=4$ field theories 
when held in a magnetic field. Nevertheless, they are still physically interesting.  
In general, extending the work of \cite{D'Hoker:2009mm}, we expect to be able to construct finite temperature black hole 
solutions which at zero temperature interpolate between $AdS_5$ in the UV\footnote{In general, in addition to the magnetic field, we expect that some of the scalar fields will also give rise to deformations of operators 
of the UV CFT.} and a given non-supersymmetric $AdS_3\times\mathbb{R}^2$ solution in the IR. The instability of the latter indicates that there will be a phase transition at finite temperature described by new types of black hole solutions and the instabilities that we discuss suggest
the types of modes that will be involved in constructing them. Some specific magnetic $AdS_3\times\mathbb{R}^2$ solutions have several different kinds of
instabilities including spatially modulated instabilities, driven by neutral scalars, and superconducting instabilities driven either
by charged scalars, charged vectors\footnote{The possibility of charged vectors producing superconducting instabilities in the presence of magnetic fields has been recently discussed in 
\cite{Ammon:2011je,Chernodub:2010qx} based on the older work of \cite{Nielsen:1978rm,Ambjorn:1979xi}. This is also reminiscent 
of ``reentrant superconductivity", reviewed in \cite{Rasolt:1992zz}, that is seen in URhGe \cite{urrhge}.}
or combinations thereof. Our results
suggests that there is a rich story involving competing phases that would be interesting to elucidate.
It is also worth emphasising that since the supersymmetric $AdS_3\times\mathbb{R}^2$ solutions are at the boundary of unstable
non-supersymmetric solutions (see figure \ref{fig:AdS3}), the corresponding dual ground states must abut different phases and hence
have the nature of quantum critical points.

We now turn to top down solutions containing $AdS_2$ factors that are supported by magnetic fields. Such solutions 
are particularly interesting since they might provide dual descriptions of locally 
quantum critical points, which have been shown to be associated with interesting non-Fermi liquid behaviour 
\cite{Lee:2008xf,Liu:2009dm,Cubrovic:2009ye,Faulkner:2009wj} (see \cite{Albash:2009wz,Basu:2009qz} for
the inclusion of magnetic fields). For $D=5$ it was shown in \cite{Almuhairi:2010rb} that the $U(1)^3$ truncation of $SO(6)$ gauged supergravity
admits a non-supersymmetric magnetic $AdS_2\times\mathbb{R}^3$ solution which can be uplifted on an $S^5$ to type IIB. 
However, it was subsequently shown in \cite{Donos:2011qt}
that, within the same truncation, this suffers from an instability involving neutral scalar fields
that are spatially modulated in the $\mathbb{R}^3$ directions. 

It has long been known that minimal $D=4$ gauged supergravity admits a non-supersymmetric magnetic
$AdS_2\times\mathbb{R}^2$ solution (the near horizon limit of the standard magnetic AdS-RN black brane solution) 
and that this can be uplifted to obtain solutions of $D=10,11$ supergravity in a variety of ways
\cite{Gauntlett:2006ai,Gauntlett:2007ma}.
The concluding section of \cite{Almuhairi:2011ws} briefly discussed how this solution can be generalised to form part of a larger family of
magnetic $AdS_2\times\mathbb{R}^2$ solutions of the $U(1)^4$ truncation of $SO(8)$ gauge supergravity which can be uplifted to $D=11$ on $S^7$. Here we will flesh out these constructions in a little more detail. 
We will find a two-parameter locus of supersymmetric solutions that includes the specific examples already mentioned
in \cite{Almuhairi:2011ws}, as special cases. 
We will also show that a large class of the non-supersymmetric solutions
suffer from similar instabilities that we find for the magnetic $AdS_3\times \mathbb{R}^2$ solutions. 
In particular we explicitly discuss an instability involving neutral scalar fields that are spatially modulated in the $\mathbb{R}^2$ directions and
a simple instability involving charged scalars.

We will also construct a supersymmetric domain wall solution that interpolates between $AdS_4$ in the UV and a representative magnetic 
$AdS_2\times\mathbb{R}^2$ solution in the IR\footnote{Note that in the context of $N=2$ $D=4$ gauged supergravity, related solutions
have been discussed in \cite{Cacciatori:2009iz,Dall'Agata:2010gj,Barisch:2011ui}.}. This solution, being supersymmetric, should describe the stable ultimate zero temperature ground state of the corresponding 
$d=3$ CFTs when they are placed in a magnetic field. This provides the first top down holographic description of a locally quantum critical point where stability is guaranteed by supersymmetry, 
and it will be interesting to investigate
the behaviour of fermion response functions for this background.

In the last part of the paper, we will briefly discuss $AdS_2$ solutions that are supported by electric fields. Indeed an
electric-magnetic duality transformation for the $U(1)^4$ truncation of $D=4$ $SO(8)$ gauged supergravity provides a simple way 
to obtain a three parameter family of electric $AdS_2\times\mathbb{R}^2$ solutions. We find that none of them preserve supersymmetry.
Furthermore, we also can use the duality transformation to find a domain wall solution, solving first order equations, that interpolates between $AdS_4$ and a non-supersymmetric
$AdS_2\times\mathbb{R}^2$ solution\footnote{In the context of $AdS_4\times SE_7$ solutions, a non-supersymmetric flow solution that interpolates between $AdS_4$ in the UV and an electric $AdS_2\times\mathbb{R}^2$ in the IR was constructed in \cite{Klebanov:2010tj} and, as yet, has not been shown to suffer from any
instabilities.}. 
We do not investigate instabilities for these solutions here,
but we expect that there will be many: for the special case of the electric $AdS_2\times\mathbb{R}^2$ solution of minimal gauged supergravity see
\cite{Denef:2009tp}.

In addition, we also construct a two parameter family of 
electric $AdS_2\times\mathbb{R}^3$ solutions of the $U(1)^3$ truncation of $D=5$ $SO(6)$ gauged supergravity. These solutions generalise
the solutions of Romans' theory given in \cite{Romans:1985ps}. Once again, none of these solutions are supersymmetric.
We will again leave a detailed analysis of instabilities to future work, but we note that it was already shown
in \cite{Donos:2011ff} that the electric $AdS_2\times\mathbb{R}^3$ solutions of Romans' theory
suffer from helical $p$-wave superconducting instabilities.

\section{Magnetic $AdS_{3}\times \mathbb{R}^{2}$ solutions}\label{magsols5}
In this section we will review the magnetic $AdS_{3}\times \mathbb{R}^{2}$ solutions of 
\cite{Romans:1985ps,Almuhairi:2010rb,Almuhairi:2011ws}. We will also
construct a supersymmetric domain wall solution that interpolates between $AdS_5$ in the
UV and the supersymmetric $AdS_{3}\times \mathbb{R}^{2}$ solution of \cite{Romans:1985ps} in the IR.

\subsection{$U(1)^3\subset SO(6)$ gauged supergravity}
We start with the $U(1)^{3}$ truncation of $D=5$ $SO(6)$ gauged supergravity \cite{Cvetic:1999xp} that keeps two neutral scalar fields
$\phi_a$. 
It is convenient to package the two scalars in terms of three constrained scalars $X_i$ via
\begin{align}
X_{1}&=e^{-\frac{1}{\sqrt{6}}\phi_{1}-\frac{1}{\sqrt{2}}\phi_{2}},\quad X_{2}=e^{-\frac{1}{\sqrt{6}}\phi_{1}+\frac{1}{\sqrt{2}}\phi_{2}},\quad X_{3}=e^{\frac{2}{\sqrt{6}}\phi_{1}}\,,
\end{align}
with $X_1X_2X_3=1$.
The Lagrangian is then given by
\begin{align}\label{eq:SO6_bh}
\mathcal{L}&=(R-V)\,\ast 1-\frac{1}{2}\,\sum_{a=1}^{2}\,\ast d\phi_{a}\wedge d\phi_{a}-\frac{1}{2}\,\sum_{i=1}^{3}(X_{i})^{-2}\,\ast F^{i}\wedge F^{i}+F^{1}\wedge F^{2}\wedge A^{3}\,,
\end{align}
where
\begin{align}
V&=-4\,\sum_{i=1}^{3}(X_{i})^{-1}\,.
\end{align}
Any solution of this theory can be uplifted on an $S^5$ to obtain an exact solution of type IIB supergravity using the formulae in \cite{Cvetic:1999xp}.

This theory can be further truncated to obtain a sector of Romans' $SU(2)\times U(1)$ gauge supergravity
theory \cite{Romans:1985ps}. This is significant because any solution of Romans' theory can also be uplifted to $D=11$ supergravity
using the general class of $M_6$ \cite{Lin:2004nb} associated with supersymmetric $AdS_5\times M_6$ solutions of $D=11$ supergravity
that are dual to $N=2$ SCFTs in $d=4$ \cite{Gauntlett:2007sm}. 
Specifically, if we set $X_1=X_2\to X$, $F^1=F^2\to F^{(3)}/\sqrt 2$, $F^3\to -G$ we obtain Romans' theory as in
\cite{Gauntlett:2007sm}, after setting the two-form to zero and identifying $F^{(3)}$ with one of the $SU(2)$ gauge-fields.
There are two other ways of obtaining Romans' theory: one by setting $X_2=X_3$, $F^2=F^3$ and another by setting
$X_1=X_3$, $F^1=F^3$. 

By setting the scalars to zero, $X_1=X_2=X_3=1$, and also $F^1=F^2=F^3$ we obtain minimal $D=5$ gauged supergravity.
Recall that any solution of this theory can be uplifted to type IIB supergravity
using the general class of $X_5$  \cite{Gauntlett:2005ww} associated with supersymmetric $AdS_5\times X_5$ solutions of type IIB supergravity
that are dual to $N=1$ SCFTs in $d=4$  \cite{Gauntlett:2006ai}
or to $D=11$ using the general class of $N_6$ \cite{Gauntlett:2004zh}  associated with supersymmetric $AdS_5\times N_6$ 
solutions of $D=11$ supergravity also dual to $N=1$ SCFTs in $d=4$  \cite{Gauntlett:2007ma}.

\subsection{The $AdS_{3}\times\mathbb{R}^{2}$ solutions}
We now consider the family of magnetic $AdS_{3}\times \mathbb{R}^{2}$ solutions to the equations of motion for
\eqref{eq:SO6_bh} found in \cite{Almuhairi:2010rb}, and studied further in \cite{Almuhairi:2011ws},
given by 
\begin{align}\label{eq:AdS2_fam}
ds_{5}^{2}&=L^{2}\,ds^{2}\left(AdS_{3}\right)+dx_{1}^{2}+dx_{2}^{2}\,,\notag\\
F^{i}&=2 q^{i}\,dx_{1}\wedge dx_{2},\quad\phi_{1}=f_{1},\quad\phi_{2}=f_{2}\,,
\end{align}
where $f_a$ are constants, 
\begin{align}\label{eq:AdS2_constants}
L^{-2}=\sum_{I=1}^3 ({\bar X_i})^{-1},\qquad 
(q^{i})^2=\bar X_i\,,
\end{align}
and $\bar X_i$ are the on-shell values
\begin{align}
\bar X_{1}&=e^{-\frac{1}{\sqrt{6}}f_{1}-\frac{1}{\sqrt{2}}f_{2}},\quad \bar X_{2}=e^{-\frac{1}{\sqrt{6}}f_{1}+\frac{1}{\sqrt{2}}f_{2}},
\quad \bar X_{3}=e^{\frac{2}{\sqrt{6}}f_{1}}\,.
\end{align}
The $q^i$ can be chosen to have either sign.
Notice that when $f_2=0$, for example, these are solutions to Romans' theory, and actually were already presented in \cite{Romans:1985ps}
and uplifted to $D=10,11$ supergravity in \cite{Gauntlett:2007sm}. When $f_1=f_2=0$ they are solutions of minimal gauged supergravity.

\begin{figure}
\centering
{\includegraphics[width=5cm]{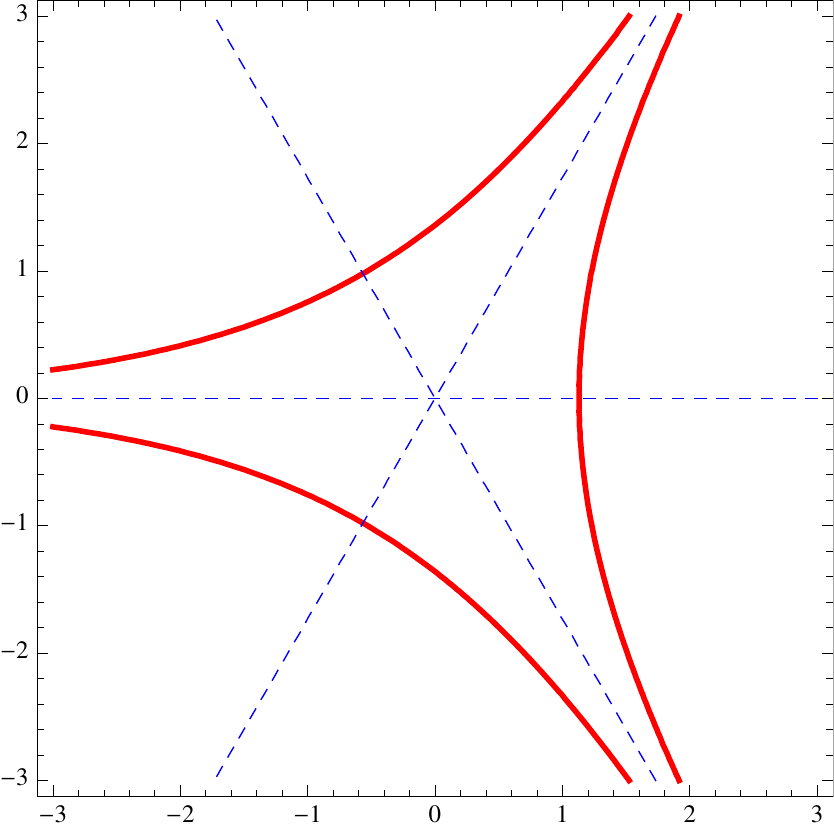}}
\caption{The moduli space of magnetic $AdS_3\times \mathbb{R}^2$ solutions. Any point in the 
$(f_1,f_2)$ plane, combined with a set of signs for the $q^i$,
gives rise to an $AdS_3\times \mathbb{R}^2$ solution. The red lines correspond to the locus of solutions that can preserve
supersymmetry, for particular choices of the signs. The dashed lines correspond to solutions that can be embedded into Romans' theory and the origin
corresponds to solutions that can be embedded in minimal gauged supergravity.}\label{fig:susy}
\end{figure}

The supersymmetry of these solutions was analysed in \cite{Almuhairi:2011ws} where it was shown that the sum of the $q^i$, with suitable signs, must vanish. 
We will review this analysis in the next subsection where we also show that the locus of supersymmetric solutions is given by
\begin{align}\label{susyconone}
2\sum_i \bar X_i^2=\left(\sum_i \bar X_i\right)^2\,.
\end{align}
We have summarised the moduli space of solutions in figure  \ref{fig:susy}.

The $AdS_3$ solutions, assuming that they are stable, are dual to $d=2$ CFTs, and the radius, $L$, 
is proportional to the central charge. We find that $L$ has a global maximum for the solution with $f_1=f_2=0$.
Along the supersymmetric branches, it is a maximum for the three solutions that can be embedded into Romans' theory
and then decreases monotonically away from them.

\subsection{Supersymmetric $AdS_5$ to $AdS_3\times\mathbb{R}^2$ domain wall}
It was already shown in \cite{Almuhairi:2011ws} that the magnetic $AdS_3\times \mathbb{R}^2$ solutions \eqref{eq:AdS2_fam},\eqref{eq:AdS2_constants} 
can be supersymmetric,
preserving two Poincar\'e supersymmetries (i.e. $(0,2)$ in $d=2$). Here we would like to show that
there are supersymmetric domain wall solutions that interpolate between 
$AdS_5$ in the UV and $AdS_3\times\mathbb{R}^2$ in the IR.
We thus consider the ansatz
\begin{align}\label{eq:radans1}
ds^{2}&=e^{2W}\,(-dt^{2}+dy^2)+d\rho^{2}+e^{2U}\,\left(dx_{1}^{2}+dx_{2}^{2} \right)\,,\notag\\
F^{i}&=2\,q^{i}\,dx_{1}\wedge dx_{2}\,,\notag\\
\phi_{a}&=\phi_{a}(\rho)\,,
\end{align}
where $W$ and $U$ are functions of $\rho$. 
We will consider the $N=1$ supersymmetry transformations as given\footnote{Note that we should set their $g=1$ and identify our $X_i$ with their $X^I=1/(3X_I)$.} in \cite{Cacciatori:2003kv}
\begin{align}\label{susyd51}
\delta\psi_{\mu}&=\nabla_{\mu}\varepsilon-\frac{i}{2} \, \sum_{i} A_{\mu}^{i}\,\varepsilon+\frac{1}{6}\,\sum_{i}X_{i}\,\gamma_{\mu}\,\varepsilon
+\frac{i}{24}\sum_{i}X_{i}^{-1}\left[\gamma_\mu{}^{\nu\rho}-4\delta^\nu_\mu \gamma^\rho\right]F^{i}_{\nu\rho}\varepsilon\,,\notag\\
\delta\lambda_{a}&=\left[-\frac{i}{4}\slashed{\partial}\phi_{a}+\frac{i}{2}\,\sum_{j}\partial_{\phi_{a}}X_{j}
+\frac{1}{8} \sum_{j}\partial_{\phi_{a}}X_{j}^{-1}{F}^{j}_{\mu\nu}\gamma^{\mu\nu}\right]\varepsilon\,.
\end{align}
To preserve these supersymmetries we require that $ \sum_{i} A_{\mu}^{i}=0$ and hence 
the magnetic charges $q^i$ should satisfy 
\begin{equation}\label{eq:cond1}
\sum_{i}q^{i}=0\,.
\end{equation}
As noted in \cite{Almuhairi:2011ws} there are another three $N=1$ supersymmetries with different sign choices for the gauge fields and hence the charges.
We expect these to correspond to the conditions $q_1+q_2-q_3=0$,
$q_1-q_2+q_3=0$ and $-q_1+q_2+q_3=0$. 
As noted in \cite{Almuhairi:2011ws}, this means that extra supersymmetry can be preserved only
if one of the charges is zero: however from \eqref{eq:AdS2_constants} we see that this is not possible,
since $X_i> 0$.

Turning now to the specific ansatz \eqref{eq:radans1}, choosing $q^{i}$ to satisfy \eqref{eq:cond1} and imposing the projection conditions
\begin{equation}\label{eq:projcond1}
\gamma_{\hat{r}}\varepsilon=-\varepsilon,\quad \gamma_{\hat{x}_{1}\hat{x}_{2}}\varepsilon=i\alpha\varepsilon,\qquad\alpha=\pm 1\,,
\end{equation}
we obtain
\begin{align}\label{eq:SUSYflow1}
-W^{\prime}+\frac{1}{3}\sum_{i}X_{i}-\frac{\alpha}{3}e^{-2U}\,\sum_{i}X_{i}^{-1}q^{i}=0\,,\notag\\
-U^{\prime}+\frac{1}{3}\sum_{i}X_{i}+\frac{2\alpha}{3}e^{-2U}\,\sum_{i}X_{i}^{-1}q^{i}=0\,,\notag\\
\phi_{a}^{\prime}+2\,\sum_{j}\,\partial_{\phi_{a}}X_{j}+2\alpha e^{-2U}\,\sum_{j}\,q^{j}\partial_{\phi_{a}}X_{j}^{-1}=0\,,\notag\\
\left[\partial_{\rho}-\frac{1}{6}\,\sum_{i}X_{i} +\frac{\alpha}{6}e^{-2U}\sum_{i}q^{i}X_{i}^{-1}\right]\varepsilon=0\,.
\end{align}
From the first and the last equation in \eqref{eq:SUSYflow1} we derive that $\varepsilon=e^{W/2}\eta$, with $\eta$ a constant spinor satisfying the projection conditions \eqref{eq:projcond1}.

For the supersymmetric $AdS_3\times \mathbb{R}^2$ solutions, we should set $W=L^{-1}\rho$, $U=0$ and $X_i=\bar X_i$ in \eqref{eq:projcond1}.
We then find the conditions
\begin{align}
L^{-1}=\frac{1}{2}\sum_i \bar X_i,\qquad
-2\alpha q^i=\bar X_i(-2\bar X_i+\sum_j \bar X_j)\,.
\end{align}
The latter condition combined with \eqref{eq:cond1} leads to the condition \eqref{susyconone}, that we mentioned earlier.
For the $AdS_5$ vacuum solution we set
$W=U=R^{-1}\,\rho$, $\phi_a=0$ and find $R=1$.

We now show that there exists supersymmetric solutions that interpolate between $AdS_5$ in the UV and a supersymmetric
$AdS_3\times\mathbb{R}^2$ solution in the IR. We do this for just one representative solution
in the IR, namely the one that exists inside Romans' theory. In fact the entire domain wall solution lies within Romans' theory
and so we set $\phi_{2}=0$ and consider the supersymmetric solution with $\phi_1=(2\sqrt{6}/3)\log 2$ and 
$q^{1}=q^{2}=-2^{-1/3}$ and $q^{3}=2^{2/3}$ (we have chosen $\alpha=1$). Within this truncation, the flow equations \eqref{eq:SUSYflow1}
for the non-trivial functions are given by
\begin{align}\label{eq:Romans_flow}
& W^{\prime}-\frac{1}{3}({2}e^{-\frac{1}{\sqrt{6}}\phi_{1}}+e^{{\frac{2}{\sqrt{6}}}\phi_{1}})
-\frac{2^{2/3}}{3}e^{-2U}\left(e^{\frac{1}{\sqrt{6}}\phi_1}-e^{-\frac{2}{\sqrt 6}\phi_{1}}\right)
=0\,,\notag\\
& U^{\prime}-\frac{1}{3}({2}e^{-\frac{1}{\sqrt{6}}\phi_{1}}+e^{{\frac{2}{\sqrt{6}}}\phi_{1}})
+\frac{2^{5/3}}{3}e^{-2U}\left(e^{\frac{1}{\sqrt{6}}\phi_1}-e^{-\frac{2}{\sqrt 6}\phi_{1}}\right)=0\,,\notag\\
& \phi_{1}^{\prime}
-\frac{4}{\sqrt 6}(e^{-\frac{1}{\sqrt{6}}\phi_{1}}-e^{{\frac{2}{\sqrt{6}}}\phi_{1}})
-\frac{2^{5/3}}{\sqrt 6}e^{-2U}\left(e^{\frac{1}{\sqrt{6}}\phi_1}+2e^{-\frac{2}{\sqrt 6}\phi_{1}}\right)
=0\,.
\end{align}

\begin{figure}
\centering
\includegraphics[width=7cm]{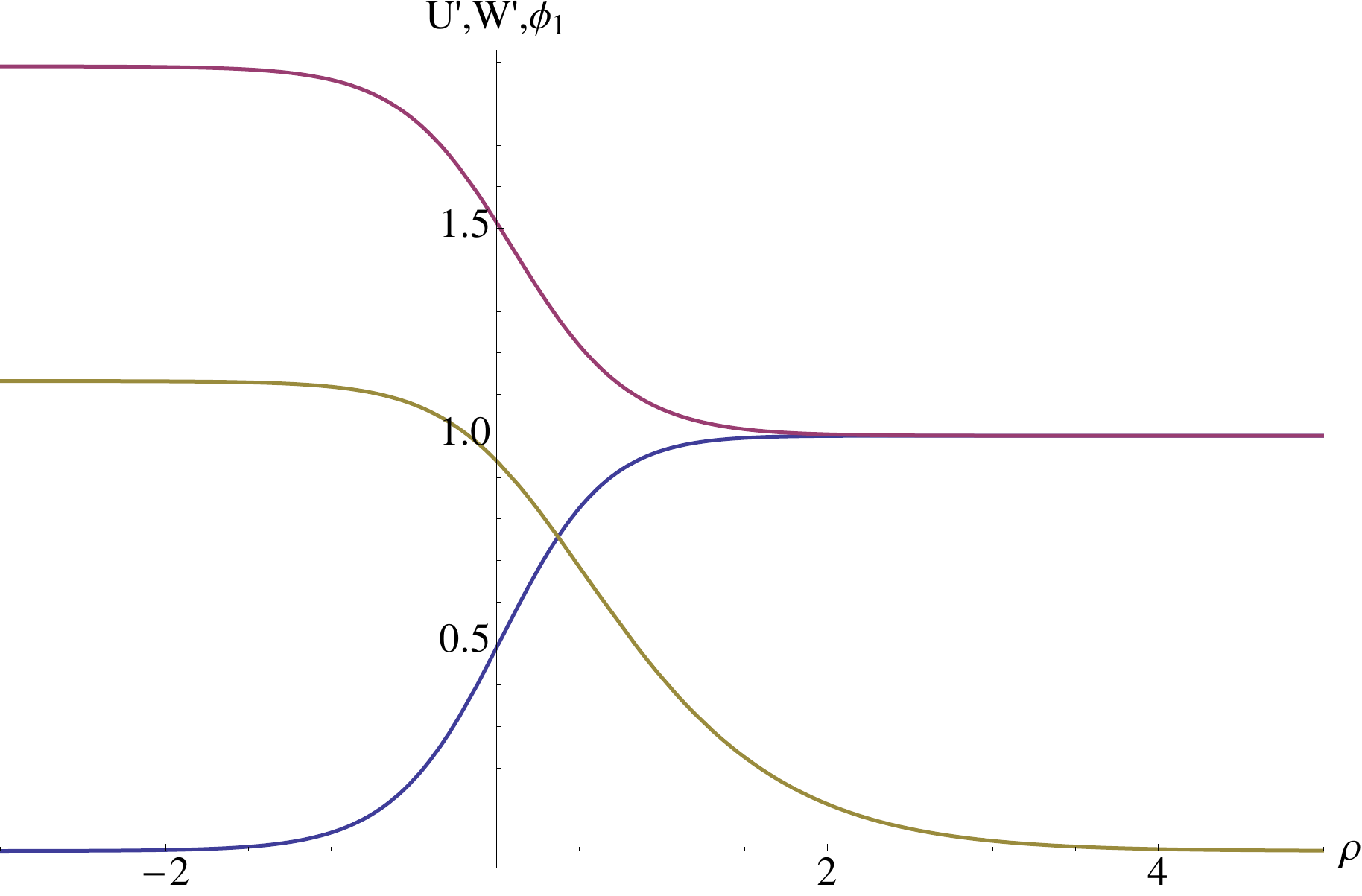}\label{fig:SU2}
\caption{We have plotted $U^{\prime}$ (blue) $W^{\prime}$ (purple) and $\phi$ (green) as functions of $\rho$ for the superssuperymmetricymmetric
domain wall solution interpolating between $AdS_{5}$ and $AdS_{3}\times\mathbb{R}^2$ that exists in Romans' theory.}
\label{fig:dwiib}
\end{figure}

Close to the $AdS_{3}\times \mathbb{R}^{2}$ solution in the far IR the system of equations \eqref{eq:Romans_flow} admits the expansion
\begin{align}\label{irbcs}
W&=w_{0}+L^{-1}\rho+\frac{1}{16}\,\left(-29+3\,\sqrt{33}\right)c_{IR}\, e^{L^{-1}\,\delta\,\rho}+\ldots\,,\notag\\
U&=c_{IR}\,e^{L^{-1}\,\delta\,\rho}+\ldots\,,\notag\\
\phi_{1}&=2\sqrt{\frac{2}{3}}\,\ln 2-\sqrt{\frac{3}{2}}\,\left(-5+\sqrt{33} \right)c_{IR}\,e^{L^{-1}\,\delta\,\rho}+\cdots\,,
\end{align}
with $w_{0}$ and $c_{IR}$ two constants while $L^{-1}=3/2^{2/3}$ and $\delta=\frac{1}{3}\left(-1+\sqrt{33} \right)$.
In the far UV we would like to approach the unit radius $AdS_{5}$ vacuum. One can easily see that the
equations \eqref{eq:Romans_flow} admits the expansion
\begin{align}\label{uvbcs}
U&=\rho+\,\mathcal{O}\left(e^{-4\rho} \right)\,,\notag\\
W&=\rho+\,\mathcal{O}\left(e^{-4\rho} \right)\,,\notag\\
\phi_{1}&=2^{7/6}\sqrt{3}\,e^{-2\rho}\rho+c_{UV}\,e^{-2\rho}+\,\mathcal{O}\left(e^{-4\rho} \right)\,,
\end{align}
where $c_{UV}$ is a constant of integration. From the above expansion we see that the scalar $\phi_{1}$, which is dual to an operator
with conformal dimension $\Delta=2$ in the UV CFT, has both a VEV and a deformation.
Using a shooting method we find that there is a solution to \eqref{eq:Romans_flow} 
with boundary conditions \eqref{irbcs} and \eqref{uvbcs} with
$w_{0}\approx-0.10$, $c_{IR}\approx 0.31$ and $c_{UV}\approx-1.97$ as we have indicated in figure \ref{fig:dwiib}.
This is the supersymmetric domain wall solution.

This solution can be uplifted on $S^5$ to type IIB or on the class of 
$M_6$ \cite{Lin:2004nb} to $D=11$ \cite{Gauntlett:2007sm}. The solutions then describe the corresponding
dual $d=4$ CFTs deformed by the presence of the magnetic field and also by the operator dual to $\phi_1$. In particular,
the uplifted supersymmetric $AdS_3\times\mathbb{R}^2$ solutions describes the IR ground state at zero temperature.

\section{Instabilities of magnetic $AdS_3\times\mathbb{R}^2$ solutions}\label{section3}
In this section we analyse various instabilities of the magnetic $AdS_3\times\mathbb{R}^2$ solutions
given in \eqref{eq:AdS2_fam},\eqref{eq:AdS2_constants} within $SO(6)$ gauged supergravity.

\subsection{Spatially modulated instabilities of the neutral scalars}\label{nscal}
Possible instabilities of the two neutral scalars appearing in the $U(1)^3$ truncation \eqref{eq:SO6_bh}
of $SO(6)$ gauge supergravity were investigated in \cite{Almuhairi:2011ws} and none were found.
In that analysis only fluctuations independent of the $x_1$ and $x_2$ direction were considered.
Here we relax this assumption and find that there are spatially modulated modes
violating the BF bound hence leading to instabilities, as summarised in figure \ref{fig:AdS3} (a).

\subsubsection{Instabilities of the solutions existing in Romans' theory}\label{s311}
Let us first discuss the instabilities of the three lines of solutions that are solutions
of Romans' theory (the dashed blue lines in figure \ref{fig:susy}). 
We should emphasise at the outset that the unstable modes that we find involve fields lying outside of Romans' theory and
hence the instability is not relevant when we uplift such solutions on $M_6$ to $D=11$
but only when we uplift them on $S^5$ to type IIB.

To illustrate we will consider the line of $AdS_3\times \mathbb{R}^2$ solutions with $f_2=0$ and $q^1=q^2$, which arise
in Romans' theory. We consider the field perturbation
\begin{align}\label{pix}
\delta A^{1}&=a(t,y,\rho) \sin(k\,x_{1})dx_2,\qquad \delta A^{2}=-a(t,y,\rho) \sin(k\,x_{1})dx_2\,,\notag\\
\delta\phi_{2}&=w(t,y,\rho)\cos(k\,x_{1})\,,
\end{align}
where $a$ and $w$ are functions of the $AdS_3$ coordinates and $k$ is a constant.
We find that the linearised equations of motion imply that
\begin{align}
(\Box_{AdS_{3}}-L^2M^2){\bf v}&=0\, ,
\end{align}
where ${\bf v}=(a,w)$, $\Box_{AdS_{3}}$ is the Laplacian of the unit radius $AdS_{3}$
and the mass matrix is given by 
\begin{align}\label{eq:mass_matrix}
M^{2}=\left(
   \begin{matrix} 
     k^{2} & 2\sqrt{2}q^1\,k \\
     4\sqrt{2}q^1(\bar X_1)^{-2}\,k & 4(\bar X_1)^{-1}+k^{2} \\
   \end{matrix}
   \right)\, .
\end{align}
Notice for $k=0$ there is no mixing at all as seen in \cite{Almuhairi:2011ws}.
After diagonalizing the mass matrix we find the two eigenvalues
\begin{equation}
m^{2}_{\pm}=\frac{1}{\bar X_1}\,\left(2+k^{2}\bar X_1\pm 2\,\sqrt{1+4k^{2}\bar X_1}\right)\,.
\end{equation}

The minimum mass is achieved on the $m_{-}^{2}$ branch when $k_{min}=\pm \frac{1}{2}\sqrt{15}(\bar X_1)^{-1/2}$ giving
\begin{equation}
m_{min}^{2}=-\frac{9}{4(2+\bar X_1^3)}\,.
\end{equation}
The $AdS_{3}$ BF bound $L^2m^{2}>-1$ is violated for $f_{1}>2\sqrt{\frac{2}{3}}\,\ln 2$. Note that when
$f_{1}=2\sqrt{\frac{2}{3}}\,\ln 2$, we have $\bar X_1=2^{-2/3}$ and the solution will satisfy the supersymmetry condition
\eqref{eq:cond1} provided that $q^3$ has opposite sign to that of $q^1=q^2$.
In other words, the supersymmetric solutions are located right at the boundary of
the set of solutions where the spatially modulated instabilities set in.
It is also worth noting that for the supersymmetric solution the static mode that saturates the BF bound, given explicitly
by
\begin{align}
w(\rho)=c_{1}e^{-\frac{\rho}{L}}\,\cos(\left|k_{min}\right|x_{1})\,,\qquad
a(\rho)=c_{2}e^{-\frac{\rho}{L}}\sin(\left|k_{min}\right|x_{1})\,,
\end{align}
where $|k_{min}|=\sqrt{15}/2^{2/3}$ and ${c_{1}^{2}}/{c_{2}^{2}}= 2^{7/3}3/5$,
preserves the supersymmetries of the background. The sign choice of $c_{1}/c_2$ 
depends on the choice of $\alpha$ in the projector \eqref{eq:projcond} and we note that for this solution the sign of $\alpha$ is
opposite to that of $q^1$.

\subsubsection{More general analysis}
We now consider perturbations about the full two parameter family of $AdS_3\times\mathbb{R}^2$ solutions
\eqref{eq:AdS2_fam},\eqref{eq:AdS2_constants}.
In general we cannot decouple the metric perturbations and so we consider 
the time independent perturbation
\begin{align}\label{eq:AdS2_pert}
\delta g_{tt}=-\delta g_{zz}&=L^{2}r^{2}\,h_{3}(r)\,\cos(k\,x_{1})\,,\notag\\
\delta g_{x_{a}x_{a}}&=h_{a}(r)\,\cos(k\,x_{1}),\quad a=1,2\,,\notag\\
\delta A^{i}&=a_{i}(r)\,\sin(k\,x_{1})dx_2,\quad i=1,2,3\,,\notag\\
\delta\phi_{a}&=w_{a}(r)\,\cos(k\,x_{1}),\quad a=1,2\,,
\end{align}
containing eight independent functions, which we take to be functions of the radial coordinate, $r$, of $AdS_3$ space when written in
Poincar\'e coordinates (with boundary located at $r\to\infty$). Expanding the equations of motion of  \eqref{eq:SO6_bh} around the solutions \eqref{eq:AdS2_fam} we find a total of eleven differential equations. After a little algebra we can show that for $k\neq 0$ the independent equations for the radial functions consist of two first order equations for $h_{1}$ and $h_{3}$ and six second order equations for each of the $a_{i}$, $w_{a}$ and $h_{2}$. Note that
the equations governing the perturbation \eqref{eq:AdS2_pert} are independent of the sign choices in \eqref{eq:AdS2_constants}.

\begin{figure}
\centering
\subfloat[]{\includegraphics[width=5cm]{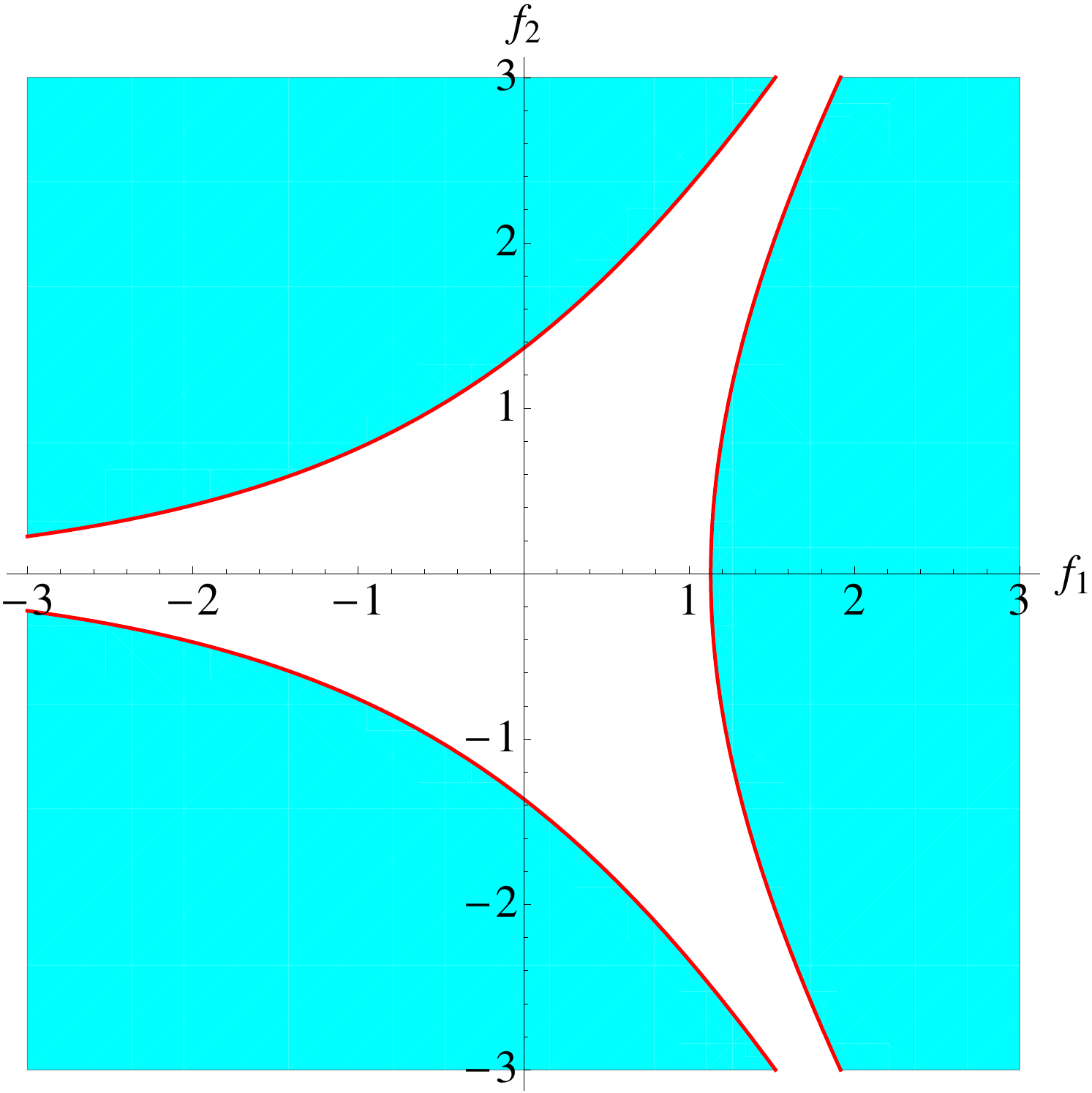}}
\subfloat[]{\includegraphics[width=5cm]{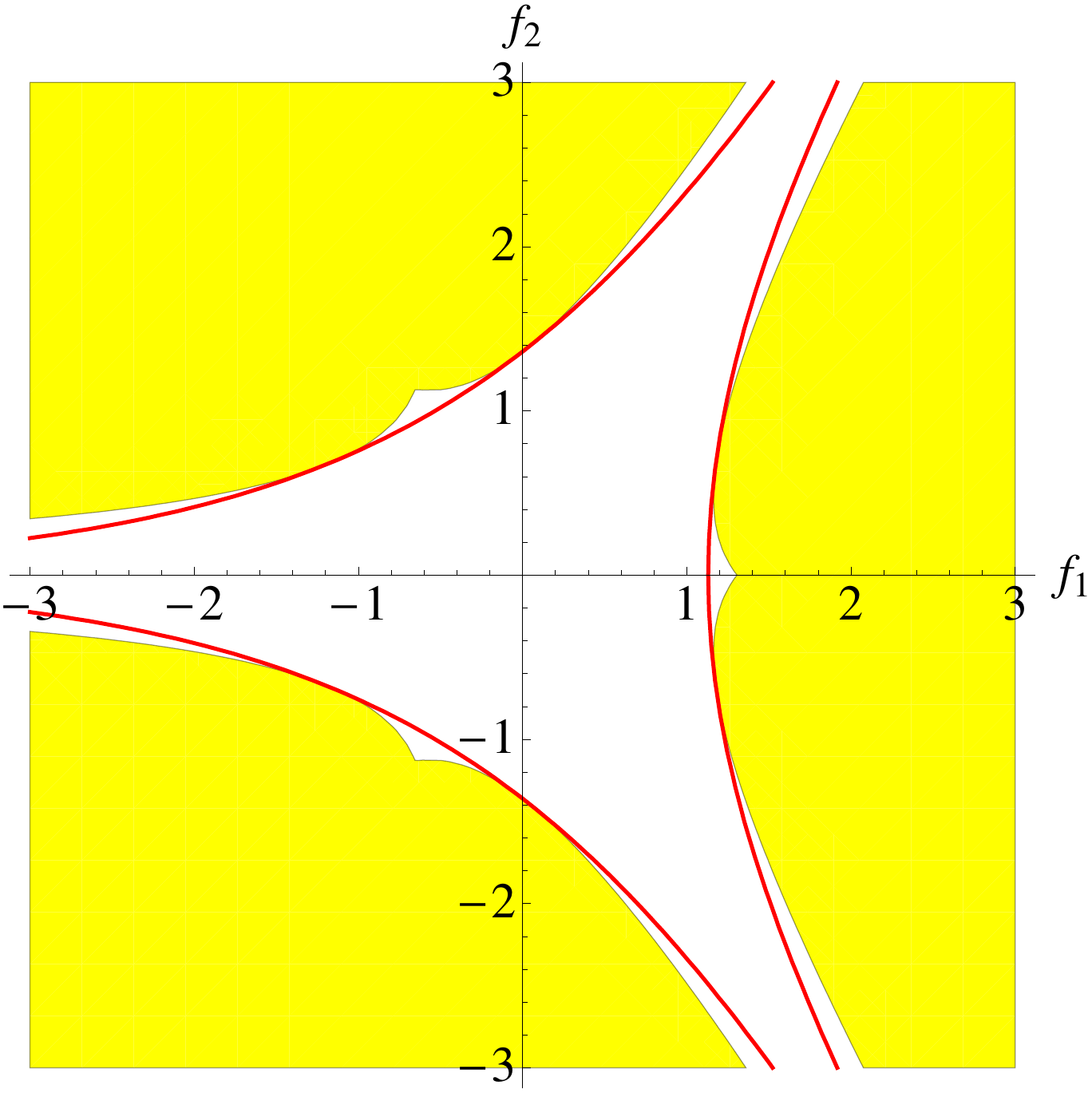}}
\subfloat[]{\includegraphics[width=5cm]{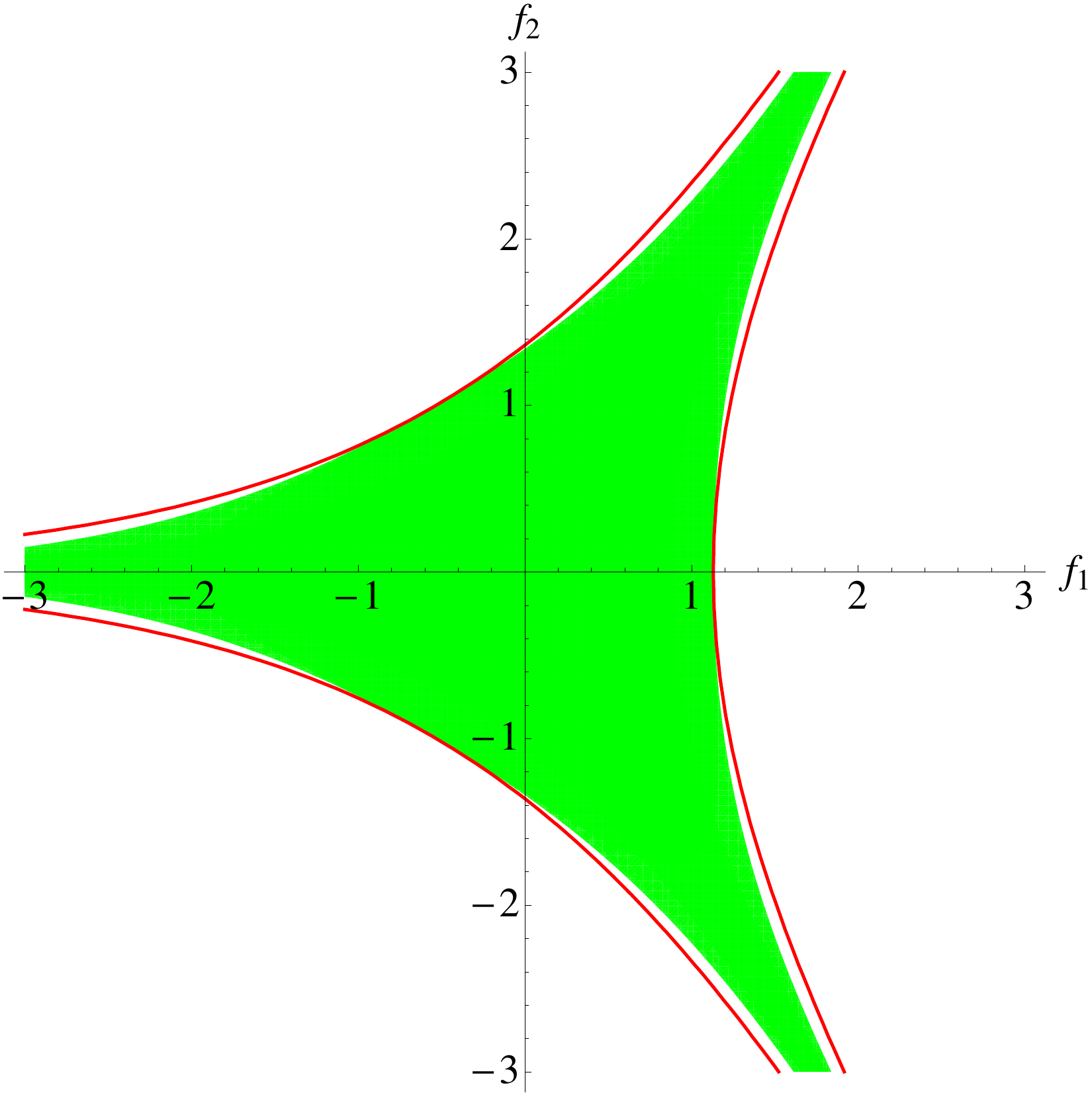}}
\caption{The shaded regions in the $(f_{1},f_{2})$ plane indicate $AdS_3\times\mathbb{R}^2$ solutions for which we find have identified
one or more unstable modes in $SO(6)$ gauged supergravity. Panel (a) indicates spatially modulated instabilities of neutral scalar fields, discussed in 
section \ref{nscal}, panel (b) indicates instabilities of the charged scalar modes discussed in section \ref{sec1p4} and panel
(c) indicates instabilities of charged vector fields discussed in section \ref{mixedcharged}. Not shown are additional instabilities of
mixed charged scalars and vectors discussed in section \ref{mixedcharged} that appear in subsets of the blue region. The supersymmetric solutions lie on the red lines.
}
\label{fig:AdS3}
\end{figure}

To find the scaling dimensions of the dual conformal field theory we look for solutions where the eight functions, as a vector,
are of the form ${\bf v}r^\delta$ where ${\bf v}$ is a constant vector and 
$\delta$ is a constant that is related to a scaling dimension in the two-dimensional conformal field theory dual to the $AdS_3  $ solution. 
The system of equations then takes the form
${\bf M v}=0$ where ${\bf M}$ is an $8\times 8$ matrix. Demanding that non-trivial values of ${\bf v}$ exist implies that $\det{\bf M}=0$
and this specifies the possible values of $\delta$ as a function of $k$. In Figure \ref{fig:AdS3} (a) we have shaded the region of the $f_{1}-f_{2}$ plane for which we find a mode with complex scaling dimension. All the unstable modes that
we find are spatially modulated with $k\neq 0$.

It is interesting to note that the boundary of the region plotted in Figure \ref{fig:AdS3} (a) (red curve) is the set of points $\left(f_{1},f_{2}\right)$ for which there exists a choice in the signs of \eqref{eq:AdS2_constants} satisfying the supersymmetry condition.
In other words, in the supersymmetric solutions there always exists a mode which saturates the $AdS_{3}$ unitarity bound at finite $k$. 
Based on the Romans' case, we expect that this mode is always supersymmetric

At the end of the last section we constructed a supersymmetric domain wall solution interpolating between $AdS_5$ in the UV
and $AdS_3\times \mathbb{R}^2$ in the IR. These solutions describe the zero temperature ground states of the dual $d=4$ CFTs when
held in a magnetic field and also deformed by the operator dual to the scalar field. The fact that the supersymmetric
$AdS_3\times\mathbb{R}^2$ solutions are at the boundary of unstable $AdS_3\times\mathbb{R}^2$ solutions indicates that
the supersymmetric ground states must adjoin different phases and hence
have the nature of quantum critical points.

\subsection{Instabilities of some charged scalars }\label{sec1p4}
We now want to consider possible instabilities of the magnetic $AdS_3\times\mathbb{R}^2$ solutions with respect to other
fields within $SO(6)$ gauged supergravity. 
We first recall that $SO(6)$ gauged supergravity has 42 scalar fields, parametrising the coset $E_{6(6)}/USp(8)$ and transforming
as ${\bf 20}+{\bf 10}+\bar{\bf 10}+{\bf 1}+{\bf 1}$ of $SO(6)$. The scalars in the ${\bf 20}$ irrep are described by
a unimodular $6\times 6$ matrix $T$ which, as we will discuss, contains the two scalars in the $U(1)^3$ truncation
\eqref{eq:SO6_bh} that we have been discussing. In this subsection and the next, we will show that there are additional instabilities\footnote{In appendix \ref{22scalars} we show that there are no instabilities in a truncation of \cite{Bobev:2010de} that
keeps four complex scalars in the ${\bf 10}+\bar{\bf 10}$ irreps (we thank N. Bobev for suggesting this calculation). We also note that the two singlet scalars comprise the
axion and dilaton; the stability of the dilaton was discussed in \cite{Almuhairi:2011ws}. We have not investigated the stability of these or the 
12 two-forms of $SO(6)$ gauged supergravity.} involving the scalar fields in $T$.
Our most general analysis will utilise the consistent truncation \cite{Cvetic:2000nc} that keeps $T$ and the 15 $SO(6)$ gauge-fields.
It is worth noting that when uplifting to type IIB, only the $D=10$ metric and five-form are involved in this truncation.

As the general analysis using the truncation \cite{Cvetic:2000nc} is rather involved we first investigate possible instabilities of three of the twenty scalars, $\varphi_i$,
appearing in $T$, using the consistent truncation of $SO(6)$ gauge supergravity \cite{Chong:2004ce}.
The Lagrangian is given by 
\begin{align}\label{eq:SO6_bhpope}
\mathcal{L}&=(R-V)\,\ast 1-\frac{1}{2}\,\sum_{a=1}^{2}\,\ast d\phi_{a}\wedge d\phi_{a}
-\frac{1}{2}\,\sum_{i=1}^{3}(X^{i})^{-2}\,\ast F^{i}\wedge F^{i}+F^{1}\wedge F^{2}\wedge A^{3}\nn
&-\frac{1}{2}\,\sum_{i=1}^{3}\,\ast d\varphi_{i}\wedge d\varphi_{i}-2\sum_{i=1}^{3}\sinh^2\varphi_i\ast A^i\wedge A^i\,,
\end{align}
with 
\begin{align}
V=&-2[ 2X^2 X^3 \cosh\varphi_2\cosh\varphi_3
 + 2X^3 X^1 \cosh\varphi_3\cosh\varphi_1
 + 2X^1 X^2 \cosh\varphi_1\cosh\varphi_2\nn
& -(X^1)^2\sinh^2\varphi_1
 -(X^2)^2\sinh^2\varphi_2
 -(X^3)^2\sinh^2\varphi_3]\,.
\end{align}

After expanding around the background \eqref{eq:AdS2_fam},
the equation of motion for the charged scalar $\varphi_{1}$ gives
\begin{equation}\label{eq:charge_eom}
\Box_{AdS_{3}}\varphi_{1}+L^{2}\,\Box_{\mathbb{R}^{2}}\varphi_{1}+4L^{2}\left(\bar X_{1}\bar X_{3}+\bar X_{1}\bar X_{2}-\bar X_{1}^{2}-(A^{1})^{2}\right)\varphi_{1}=0\,,
\end{equation}
where, again, $\Box_{AdS_{3}}$ is the Laplacian of the unit radius $AdS_{3}$.
The equations for $\varphi_{2,3}$ are given by cyclic permutations of indices. We now choose a gauge such that $A^{i}=q^{i}\,\left(x_{1}\,dx_{2}-x_{2}\,dx_{1}\right)$ while for the scalar we consider the lowest Landau level ``ground state'' 
\begin{equation}
\varphi_{1}=e^{-|q_1|\,\left(x_{1}^{2}+x_{2}^{2}\right)}\,\psi_{1}(t,r,z)\,,
\end{equation}
 giving
 \begin{equation}
(\Box_{AdS_{3}}-L^{2}m_{\psi_{1}}^{2})\psi_{1}=0\,,
\end{equation}
where
\begin{equation}
m_{\psi_{1}}^{2}=-4\,\left(\bar X_{1}\bar X_{3}+\bar X_{1}\bar X_{2}-\bar X_{1}^{2}-|q_1| \right)\,,
\end{equation}
which agrees with the last line of equation (6.7) in \cite{Almuhairi:2011ws} (after setting their $g=1$).

In figure \ref{fig:AdS3} (b) we have indicated where these modes violate the BF bound.
As one can see from figure \ref{fig:AdS3} (b), these modes intersect the locus of supersymmetric solutions
in six places. Using the results of \cite{Liu:2007rv} (see eq. (2.24) and set $g=1$) one
can check that at these points the modes saturating the BF bound are supersymmetric, just as we saw in section \ref{s311}.
The higher Landau levels have mass
 \begin{equation}
m_{\psi_{1}}^{2}=-4\,\left(\bar X_{1}\bar X_{3}+\bar X_{1}\bar X_{2}-\bar X_{1}^{2}-|q_1|(2n+1) \right)\,,
\end{equation}
again as in \cite{Almuhairi:2011ws}. These are unstable in sub-regions of figure \ref{fig:AdS3} (b), and in particular
do not intersect the supersymmetric locus.

It is interesting to note that all of these instabilities involve electrically charged fields and hence are
associated with new branches of finite temperature superconducting black brane solutions. As the superconductivity is being driven
by a magnetic field, it would be interesting
to construct and study them further.

\subsection{Instabilities of charged scalars and vectors}\label{mixedcharged}

We will now examine perturbations of $SO(6)$ gauged SUGRA contained in the truncation \cite{Cvetic:2000nc}. 
This contraction contains twenty scalar fields arranged in a unimodular, $6\times 6$ symmetric matrix $T_{ij}$ and keeps
all of the $SO(6)$ gauge fields.
The vector and scalar equation of motion we would like to perturbatively expand are
\begin{align} \label{eq:vectoreq}
&-\,D\ast DT_{ij}+T^{-1}{}^{k_{1}k_{2}}\,DT_{ik_{1}}\wedge\ast DT_{k_{2}j}=-2\left(T_{ik_{1}}T^{k_{1}k_{2}}T_{k_{2}j}-T_{ik_{1}}T^{k_{1}}{}_{j}\,T_{k}{}^{k} \right)\notag\\
&+T^{-1}_{lm}\,\ast F^{l}{}_{i}\wedge F^{m}{}_{j}-\frac{1}{6}T_{ij}\,\left[-2\,\left(T_{lk}T^{lk}-\left(T_{k}{}^{k} \right)^{2} \right)+T^{-1}_{pk}T^{-1}_{lm}\,\ast F^{lk}\wedge F^{mp} \right]\,,\nn
&\qquad\qquad\qquad \qquad D\left(T^{-1}_{ik}T^{-1}_{jl}\ast F^{kl} \right)=-2\,T^{-1}_{k\left[i\right.}\,\ast DT_{\left. j\right]}{}^{k}\,,
\end{align}
where 
\begin{align}
DT_{ij}&=dT_{ij}+A_{i}{}^{k}T_{kj}-T_{ik}A^{k}{}_{j} \,,\nn
F_{ij}&=dA_{ij}+A_{ik}\wedge A^{k}{}_{j}\,,
\end{align}
and we note that the first line of \eqref{eq:vectoreq} corrects a sign in \cite{Cvetic:2000nc}.

We find it convenient to switch to a complex notation which just keeps the $SU(3)\subset SO(6)$ symmetry manifest.
We will write the magnetic $AdS_{3}\times\mathbb{R}^{2}$ solutions \eqref{eq:AdS2_fam},\eqref{eq:AdS2_constants} as
\begin{align}\label{bgroundtext}
\tilde{T}_{I\bar{I}}=X_{I},\qquad
\tilde{A}_{I\bar{I}}=q_I\left(\bar{z}\,dz-z\,d\bar{z} \right)\,,
\end{align}
where $I,J=1,\dots 3$, are $SU(3)$ indices and
$z=\frac{1}{\sqrt{2}}\,\left(x^{1}+i\,x^{2} \right)$. Note that $X_I$ are the on-shell values, $X_I=\bar X_i$, and also $q_I=q^i$ (hence $q_I=\pm X_I^{1/2}$). 
Consider the scalar perturbation $T=\tilde{T}+t$
where $t$ is a complex matrix. The perturbations $t_{I\bar I}$ correspond to perturbations of the neutral scalar
fields that we considered in section \ref{sec1p4}, while the perturbations $t_{II}$ correspond to the charged scalars\footnote{Note that
there are three more scalar fields complementing the real $\varphi_i$ that we didn't explicitly consider.}
 that we considered
in section \ref{sec1p4}. Thus we now consider perturbations $t_{IJ}$ and $t_{I\bar J}$ with $I\neq J$. We note that
these modes were considered in \cite{Almuhairi:2011ws} but the mixing between these modes and the charged modes in the gauge fields
was overlooked and we will obtain different results for the spectrum.

We thus consider
\begin{align}
 T=\tilde{T}+t,\qquad A=\tilde A +a\,,
\end{align}
where we are expanding around the background \eqref{bgroundtext}. Furthermore, we find that it is consistent to set the components of 
the one-form $a$ along the $AdS_3$ directions to vanish and so we write
\begin{equation}
a=a^{1}\,dz+a^{2}\,d\bar{z}\,,
\end{equation}
where $a^{i}$ and $t$ are complex matrices that are functions of both the $AdS_{3}$ coordinates and also $\left(z,\bar{z} \right)$.

We will first consider the modes $a^i_{IJ}$, $t_{IJ}$, $I\ne J$. Linearising the equations of motion and introducing an appropriate set 
of ladder operators we are led to the following spectrum (see appendix \ref{appendixcharged} for details). Here
we define
\begin{align}
\omega_{IJ}=q_I+q_J,\qquad
W_{IJ}=sign({\omega_{IJ}})(q_I^{-1}-q_J^{-1}),\qquad
V_{IJ}=X_{I}-X_{J}\,.
\end{align}
(note that $W$ is defined slightly differently in the appendix).
When $\omega_{IJ}\ne0$, the independent modes are labelled by two integers $n,m>0$.
There is a tower of modes with $n=0$, just involving the charged vector fields $a^i_{IJ}$, which have $AdS_3$ mass given by
\begin{equation}\label{zms}
m_{0,m}^{2}=-2|\omega_{IJ}|-2W_{IJ}V_{JI}+V^2_{IJ}\,.
\end{equation}
Also, for each $m$ and $n$ there are mixed modes, involving both $a^i_{IJ}$ and $t_{IJ}$, with mass matrix
\begin{align}\label{eq:mass_matrix2t}
M^{2}_{n,m}=\left(
   \begin{matrix} 
     2|\omega_{IJ}|\left(2n+1\right)-4q^{-1}_Iq_J^{-1}+2W_{IJ}V_{JI} +V_{IJ}^{2}& 4\sqrt{2|\omega_{IJ}|}W_{IJ} \left(n+1\right)^{1/2}\\
     2\sqrt{2|\omega_{IJ}|}W_{IJ}\,\left(n+1\right)^{1/2} & 2|\omega_{IJ}|\left(2n+1\right)-2W_{IJ}V_{JI}+V_{IJ}^{2} \\
   \end{matrix}
   \right)\, .
\end{align}
In figure \ref{fig:AdS3} (c) we have indicated where the zero modes \eqref{zms} can violate the $AdS_3$ BF bound leading to an instability. The modes arising from diagonalising \eqref{eq:mass_matrix2t} can also violate
the BF bound, but only in the region outside of the three supersymmetry lines (in particular there is no overlap with the zero mode instabilities 
in figure \ref{fig:AdS3} (c)). It is straightforward to determine, numerically, which of the
diagonalised modes has the largest violation of the BF bound. The fact that $n$ is as integer leads to a more elaborate structure as compared to the spatially modulated modes labelled by a continuous variable in figure \ref{fig:AdS3} (a).

When $\omega_{I J}=0$, which occurs along the three lines with $X_I=X_J$ (the dashed lines in figure \ref{fig:susy}), 
we are essentially led back to the mass matrix that we saw 
for the spatially modulated neutral scalars in \eqref{eq:mass_matrix}.

The story for the modes $a^i_{I\bar J}$, $t_{I\bar J}$, with $I\ne J$ is very similar. We now define
\begin{align}
\omega_{I\bar J}=q_I-q_J,\qquad
W_{I\bar J}=sign(\omega_{I\bar J})(q_I^{-1}+q_J^{-1}).
\end{align}
When $\omega_{I\bar J}\ne0$,
the independent modes are again labelled by two integers $n,m>0$.
Again there is a tower of modes with $n=0$, again just involving the vector fields, with $AdS_3$ mass given by
\begin{equation}\label{zms2}
m_{0,m}^{2}=-2|\omega_{I\bar J}|-2W_{I\bar J}V_{JI}+V^2_{IJ}\,.
\end{equation}
Then for each $m$ and $n$ there are mixed modes with mass matrix
\begin{align}\label{eq:mass_matrix21}
M^{2}_{n,m}=\left(
   \begin{matrix} 
     2|\omega_{I\bar J}|\left(2n+1\right)+4q^{-1}_Iq_J^{-1}+2W_{I\bar J}V_{JI} +V_{IJ}^{2}& 4\sqrt{2|\omega_{I\bar J}|}W_{IJ} \left(n+1\right)^{1/2}\\
     2\sqrt{2|\omega_{I\bar J}|}W_{IJ}\,\left(n+1\right)^{1/2} & 2|\omega_{I\bar J}|\left(2n+1\right)-2W_{I\bar J}V_{JI}+V_{IJ}^{2} \\
   \end{matrix}
   \right)\, .
\end{align}
The instabilities are similar to those we saw above. In particular, the zero modes 
\eqref{zms2} can violate the $AdS_3$ BF bound leading to an instability as indicated
in figure \ref{fig:AdS3} (c)), and there can also be BF violating modes in \eqref{eq:mass_matrix21}. 
When $\omega_{I \bar J}=0$ the situation is again analogous to the
spatially modulated neutral scalars in \eqref{eq:mass_matrix}.

Note that the instabilities of these charged modes, mixing vectors and scalars, are again associated with holographic superconductivity. These instabilities
are somewhat similar to those involving only charged gauge fields in the presence of a magnetic field that were studied in 
\cite{Ammon:2011je}, building on \cite{Chernodub:2010qx,Nielsen:1978rm,Ambjorn:1979xi}.
We also note that there are various modes which we have studied in this section that saturate the BF bound and intersect with the locus of supersymmetric solutions. We expect them to preserve the supersymmetry but we have not checked the details. 

\subsection{Discussion}
Figure \ref{fig:AdS3} summarises most\footnote{It does not include the instabilities arising from
\eqref{eq:mass_matrix2t} and \eqref{eq:mass_matrix21} which, as we discussed, lie in a subset of the cyan coloured regions.}
 of the instabilities that we have found within $SO(6)$ gauge supergravity which 
are thus relevant to $N=4$ SYM theory after uplifting on $S^5$. We see that
apart from the supersymmetric solutions, there is only a very small range of parameters for which
we have not found an instability. It would be interesting to know whether or not those solutions are in fact stable within type IIB 
supergravity. The general picture that has emerged here and in \cite{Almuhairi:2011ws} is that studying $N=4$ SYM in a magnetic field using holography is not a straightforward proposition.

It is worth discussing which of the instabilities that we have discussed reside within the truncation to Romans' theory.
This is relevant if we uplift the solutions not on $S^5$ to type IIB but on the general class $M_6$ of \cite{Lin:2004nb} to $D=11$ \cite{Gauntlett:2007sm}.
To be specific we consider the truncation $\phi_2=0$, i.e. $X_1=X_2\equiv X$ and $A^1=A^2$. 
As we already mentioned there is no longer spatially modulated instabilities of the neutral
scalar $X$. In the language of this section, the three $SU(2)$ gauge fields of Romans' theory
can be identified with the real $A_{1\bar 1}=A_{2\bar 2}$ and the complex $A_{12}$. Recall that Romans'
theory does not have charged scalar fields. Putting this together, we find that instabilities only arise in \eqref{zms} with $IJ=12$ after noting that 
$q^1=q^2$, $W_{12}=V_{12}=0$. Indeed
we find instabilities for the solutions in the range $-2.00\lesssim  f_1\lesssim 0.87$ (recall that the supersymmetric
solution has $f_1\sim 1.13$).

Finally, we note that none of the instabilities that we have discussed appear in minimal gauged supergravity. This is relevant when
we uplift the single magnetic $AdS_3\times \mathbb{R}^2$ solution on the general class $X_5$ \cite{Gauntlett:2005ww} 
to type IIB supergravity \cite{Gauntlett:2006ai} or on the general class of $N_6$ \cite{Gauntlett:2004zh}  to
 $D=11$ supergravity \cite{Gauntlett:2007ma}.

\section{Magnetic $AdS_{2}\times \mathbb{R}^{2}$ solutions}\label{mag11}
In this section we construct magnetic $AdS_{2}\times \mathbb{R}^{2}$ solutions
of $D=4$ $SO(8)$ gauged supergravity, which are analogous to the $D=5$ solutions of section \ref{magsols5}.
We also construct a supersymmetric domain wall solution that interpolates between $AdS_4$ in the UV
and a supersymmetric $AdS_{2}\times \mathbb{R}^{2}$ solution in the IR.

\subsection{$U(1)^{4}\subset SO(8)$ gauged supergravity}

We consider the $U(1)^{4}$ truncation of $D=4$ $SO(8)$ gauged supergravity that keeps
three neutral scalar fields $\phi_a$ \cite{Cvetic:1999xp}. The Lagrangian is given by 
\begin{align}\label{eq:4dLag}
\mathcal{L}&=\frac{1}{2}R-\frac{1}{4}\,\sum_{a=1}^{3}\left(\partial\phi_{a} \right)^{2}-\sum_{i=1}^{4} X_{i}^{-2}\,\left(F^{i}\right)_{\mu\nu}\left(F^{i}\right)^{\mu\nu}-V\left(X_{i} \right)\,,
\end{align}
where
\begin{align}
&X_{1}=e^{\frac{1}{2}\left(-\phi_{1}-\phi_{2}-\phi_{3} \right)},\quad X_{2}=e^{\frac{1}{2}\left(-\phi_{1}+\phi_{2}+\phi_{3} \right)},\quad X_{3}=e^{\frac{1}{2}\left(\phi_{1}-\phi_{2}+\phi_{3} \right)},\quad X_{4}=e^{\frac{1}{2}\left(\phi_{1}+\phi_{2}-\phi_{3} \right)}\,,\notag\\
&V\left(X_{i} \right)=-\frac{1}{2}\,\sum_{i\neq j}\,X_{i}X_{j}=-2\left(\cosh\phi_1+\cosh\phi_2+\cosh\phi_3\right)\,,
\end{align}
and we note $X_1 X_2 X_3 X_4=1$.
Any solution of this theory that satisfies $F^{i}\wedge F^{j}=0$ can be uplifted\footnote{
To do this we should set $g^2=1/2$ in eq. (3.8) of \cite{Cvetic:1999xp}
and identify $(F^{i})^{there}=2\sqrt{2}(F^{i})^{here}$. It is also worth noting that we are using the same conventions as in
\cite{Duff:1999gh} setting $g=1$ there.}
to $D=11$ using the formulae in \cite{Cvetic:1999xp};  all of the solutions and the linearised modes that we consider in this section satisfy this condition.

Note that it is consistent to further truncate by setting
$X_2=X_3=X_4$ along with $F^2=F^3=F^4$ to obtain a sector of the $SU(3)$ invariant subsector of $SO(8)$
gauged supergravity \cite{Warner:1983vz}\cite{Bobev:2010ib} and the corresponding uplifted solutions will have $SU(3)\times U(1)^2$ symmetry.
This is a case that we will sometimes focus on in the sequel. Alternatively it is also consistent to set $X_1=X_2$, $X_3=X_4$ along with 
$F^1=F^2$ and  $F^3=F^4$ and the corresponding uplifted solutions will have $SU(2)^2\times U(1)^2$ symmetry.
Both of these theories can be further truncated to minimal gauged supergravity by setting all of the scalars to zero,
$X_1=X_2=X_3=X_4=1$, and $F^1=F^2=F^3=F^4$. Solutions of minimal $D=4$ gauged theory can be uplifted to $D=10$ and $D=11$
using manifolds associated with general classes of $AdS_4\times M_7$ solutions dual to $N=2$ SCFTs in 
$d=3$ \cite{Gauntlett:2007ma}, including $SE_7$ and also those of section 7.2 of  \cite{Gauntlett:2006ux}.
 
We now look for the most general class of $AdS_{2}\times \mathbb{R}^{2}$ solutions to the equations of motion of
\eqref{eq:4dLag} that are supported by magnetic fluxes. We thus consider
\begin{align}\label{eq:AdS2ansatz}
ds^{2}&=L^{2}\,ds^{2}\left(AdS_{2}\right)+dx_{1}^{2}+dx_{2}^{2}\,,\notag\\
F^{i}&=\tfrac{1}{2}q^{i}\,dx_{1}\wedge dx_{2}\,,\nn
\phi_1&=f_1,\quad \phi_2=f_2,\quad\phi_3=f_3\,,
\end{align}
where $q^i$, $f_a$ are constants and $L$ is the $AdS_2$ radius.
If we define the on-shell quantities
\begin{align}\label{exs}
\bar X_{1}&=e^{\frac{1}{2}\left(-f_{1}-f_{2}-f_{3} \right)},\quad \bar X_{2}=e^{\frac{1}{2}\left(-f_{1}+f_{2}+f_{3} \right)},\quad \bar X_{3}=e^{\frac{1}{2}\left(f_{1}-f_{2}+f_{3} \right)},\quad \bar X_{4}=e^{\frac{1}{2}\left(f_{1}+f_{2}-f_{3} \right)}\,,
\end{align}
we find that there is a three parameter family of solutions specified by arbitrary values of $(f_1,f_2,f_3)$ with 
\begin{align}\label{eq:ads2sol}
\left(q^{i}\right)^{2}=\frac{\bar{X}_{i}^{2}}{2}\,\sum_{j\neq k\neq i} \bar{X}_{j}\bar{X}_{k},\qquad
L^{-2}=-2\, V\left(\bar{X}_{i} \right)\,,
\end{align}
and we note that the $q^{i}$ can be chosen to have either sign.

For the $SU(3)\times U(1)^2$ symmetric subspace of solutions we take $f_1=f_2=f_3$ and $q^2=q^3=q^4$.
This gives rise to a one-dimensional family of solutions labelled by $\bar{X}\equiv \bar{X}_{2}=\bar{X}_{3}=\bar{X}_{4}=(\bar X_1)^{-1/3}$
with
\begin{align}\label{su3invt}
(q^1)^2=\frac{3}{\bar X^4},\qquad
(q^2)^2=(2+\bar X^4),\qquad
L^2=\frac{\bar X^2}{6(1+\bar X^4)}\,.
\end{align}

\subsection{SUSY fixed points}
We next want to investigate which of these magnetic $AdS_2\times\mathbb{R}^2$ 
solutions are supersymmetric. In the next subsection we will also investigate the possibility of
supersymmetric flows that interpolate between $AdS_4$ in the UV and $AdS_2\times\mathbb{R}^2$ in the IR.
We thus consider the ansatz
\begin{align}\label{eq:radans}
ds^{2}&=-e^{2W}\,dt^{2}+d\rho^{2}+e^{2U}\,\left(dx_{1}^{2}+dx_{2}^{2} \right)\,,\notag\\
F^{i}&=\tfrac{1}{2}q^{i}\,dx_{1}\wedge dx_{2}\,,\notag\\
\phi_{a}&=\phi_{a}(\rho)\,.
\end{align}
where $W,U$ are functions of $\rho$.

The supersymmetry variations for the $U(1)^4$ truncation of $SO(8)$ gauged supergravity were analysed in \cite{Duff:1999gh}
and it was shown that it is convenient to break up the $\mathcal{N}=8$ real Killing spinors into four pairs. We would like 
to preserve some of the Poincar\'e supersymmetries of $AdS_{4}$ and a consideration of equations (2.15)-(2.16) in \cite{Duff:1999gh} implies that depending on which of the four pairs of supersymmetries that we want to preserve we should impose one of the conditions
\begin{align}\label{omeq}
q^1+q^2+q^3+q^4=0, \qquad q^1+q^2-q^3-q^4=0,\nn q^1-q^2+q^3-q^4=0,\qquad  q^1-q^2-q^3+q^4=0.
\end{align}
For definiteness, let us choose to preserve the supersymmetries corresponding to
\begin{equation}\label{eq:cond}
\sum_{i}q^{i}=0\,.
\end{equation}
The associated supersymmetry variations were written down in (4.3) of \cite{Duff:1999gh}. After suitably comparing our notation with that
of \cite{Duff:1999gh}, and
switching from two real spinor parameters to a complex spinor parameter $\epsilon$, we have
\begin{align}\label{gensusvar2}
\frac{1}{2}\,\delta\psi_{\mu}&=\nabla_{\mu}\varepsilon+i \, \sum_{i} A_{\mu}^{i}\,\varepsilon+\frac{1}{4\sqrt{2}}\,\sum_{i}X_{i}\,\gamma_{\mu}\,\varepsilon-i\frac{1}{4\sqrt{2}}\sum_{i}X_{i}^{-1}\slashed F^{i}\gamma_{\mu}\varepsilon\,,\notag\\
2\,\delta\chi_{a}&=\left[i\sqrt{2}\,\slashed{\partial}\phi_{a}-i 2\,\sum_{j}\partial_{\phi_{a}}X_{j}+2 \sum_{j}\partial_{\phi_{a}}X_{j}^{-1}\slashed{F}^{j}\right]\varepsilon\,.
\end{align}

Turning now to the specific ansatz \eqref{eq:radans}, choosing $q^{i}$ to satisfy \eqref{eq:cond} and imposing the projection conditions
\begin{equation}\label{eq:projcond}
\gamma_{\hat{r}}\varepsilon=-\varepsilon,\quad \gamma_{\hat{x}_{1}\hat{x}_{2}}\varepsilon=i\alpha\varepsilon,\qquad\alpha=\pm 1\,,
\end{equation}
we obtain 
\begin{align}\label{eq:SUSYflow}
-W^{\prime}+\frac{1}{2\sqrt{2}}\sum_{i}X_{i}+\frac{\alpha}{2\sqrt{2}}e^{-2U}\,\sum_{i}X_{i}^{-1}q^{i}=0\,,\notag\\
-U^{\prime}+\frac{1}{2\sqrt{2}}\sum_{i}X_{i}-\frac{\alpha}{2\sqrt{2}}e^{-2U}\,\sum_{i}X_{i}^{-1}q^{i}=0\,,\notag\\
-\sqrt{2}\phi_{a}^{\prime}-2\,\sum_{j}\,\partial_{\phi_{a}}X_{j}+2\alpha e^{-2U}\,\sum_{j}\,q_{j}\partial_{\phi_{a}}X_{j}^{-1}=0\,,\notag\\
\left[\partial_{\rho}-\frac{1}{4\sqrt{2}}\,\sum_{i}X_{i} -\frac{\alpha}{4\sqrt{2}}e^{-2U}\sum_{i}q^{i}X_{i}^{-1}\right]\varepsilon=0\,.
\end{align}
From the first and the last equation in \eqref{eq:SUSYflow} we derive that $\varepsilon=e^{W/2}\eta$ with $\eta$ a constant spinor satisfying the projection conditions \eqref{eq:projcond}.

To determine which of the $AdS_{2}\times\mathbb{R}^{2}$ solutions, summarised in 
\eqref{eq:AdS2ansatz}, \eqref{exs}, \eqref{eq:ads2sol}, are supersymmetric, we 
set $W=L^{-1}\rho$, $U=0$ and $X_i=\bar X_i$ in \eqref{eq:projcond}.
We find that the magnetic charges are given by
\begin{align}\label{expqs}
2\alpha q_i=\bar X_i(-2\bar X_i+\sum_j \bar X_j)\,,
\end{align}
and \eqref{eq:cond} then gives the condition
\begin{align}\label{susyconq}
2\sum_i \bar X_i^2=\left(\sum_i \bar X_i\right)^2\,,
\end{align}
analogous to what we saw in the $D=5$ case \eqref{susyconone}.
One can directly check that the conditions in \eqref{eq:ads2sol} are satisfied (as expected)
and that the radius of the $AdS_2$ factor can now also be written
\begin{equation}
L^{-1}=\frac{1}{\sqrt{2}}\,\sum_{i}\bar{X}_{i}\,.
\end{equation}

At this point we have shown that any solutions to the flow equations \eqref{eq:SUSYflow} 
preserve 1/16 of the supersymmetries, i.e. two Poincar\'e supersymmetries, which is enhanced to
1/8 supersymmetry for the $AdS_2$ fixed points. For there to be supersymmetry enhancement, one needs to
have solutions to another of the conditions in \eqref{omeq}, but this is not compatible with \eqref{expqs}, \eqref{susyconq}.

\begin{figure}
\centering
\includegraphics[width=0.4\textwidth]{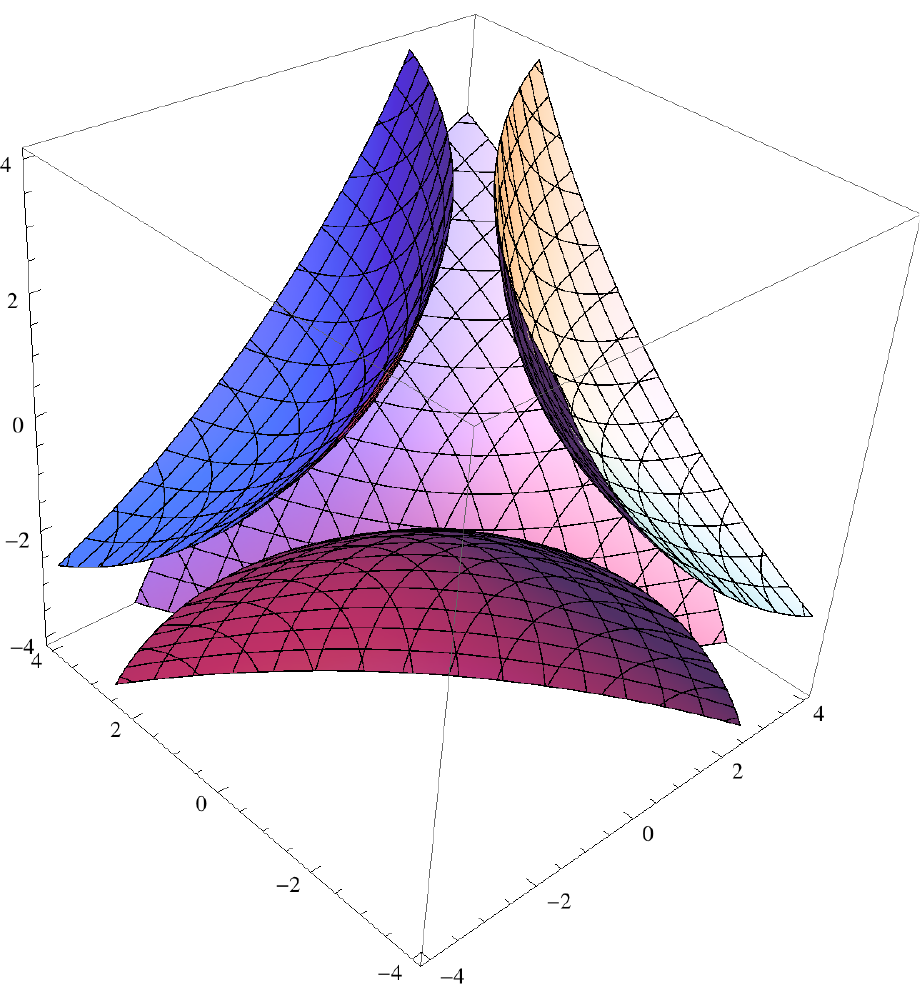}
\caption{The moduli space of supersymmetric magnetic $AdS_2\times \mathbb{R}^2$ solutions in the $f_1,f_2,f_3$ space.}\label{3d}
\end{figure}

In the three-dimensional moduli space of solutions, labelled by $(f_1,f_2,f_3)$, we have a two dimensional locus of supersymmetric solutions
fixed by \eqref{susyconq}, which we have plotted in figure \ref{3d}. Let us discuss a few special cases. Firstly, there are supersymmetric solutions when one of the $f_a$ is set to zero.
However, there are no supersymmetric solutions when two of the $f_a$ are set to zero. In particular, there are no supersymmetric solutions
with $U(1)^2\times SU(2)^2$ symmetry that have e.g. $X_1=X_2$, $X_3=X_4$, $q^1=q^2$ and $q^3=q^4$, as noted in
\cite{Almuhairi:2011ws}. The $AdS_2$ solutions of minimal gauged supergravity with all $f_a$ zero are not supersymmetric, as is well known.

Secondly, there are supersymmetric solutions when we set two of the $f_a$ equal. Furthermore, there is a single supersymmetric solution
when we set all of them to be equal, $f_1=f_2=f_3$. Specifically, in the $SU(3)\times U(1)^2$ invariant class solutions given in
\eqref{su3invt} we should take $q^1=-3 q^2$ (a condition that was also noted in \cite{Almuhairi:2011ws}) with
\begin{align}\label{eq:SU3fp}
\bar{X}=\left(-1+\frac{2}{\sqrt{3}}\right)^{1/4},\quad  q^2=\frac{1}{3\alpha}\,\sqrt{9+6\sqrt{3}},\quad
L^{-1}=2\,\left(9+6\sqrt{3}\right)^{1/4}\,.\end{align}
We show in appendix \ref{dircon} that the uplifted $D=11$ metric for this solution, using the formulae in \cite{Cvetic:1999xp}, can be recast in the
formalism of \cite{Kim:2006qu}. This provides a direct and very satisfying check on the supersymmetry of the solution.

\subsection{Supersymmetric $AdS_4$ to $AdS_2\times\mathbb{R}^2$ domain wall}\label{sdw4}

We will be interested in constructing a supersymmetric domain wall solution that describes a flow from the $AdS_4$ vacuum
to the fixed point \eqref{eq:SU3fp} which preserves $U(1)^{2}\times SU(3)$ in eleven dimensions. 
To construct the flow we truncate to the $SU(3)$ invariant sector by setting $\phi_{a}=\phi$ in \eqref{eq:SUSYflow} to obtain the first order system of equations
\begin{align}\label{eq:SU3flow}
W^{\prime}-\frac{1}{2\sqrt{2}}e^{-2U-\frac{3}{2}\phi}\,
\left(e^{2U}(1+3 e^{2\phi})+\sqrt{9+6\sqrt{3}}(e^{\phi}-e^{3\phi}) \right)=0\,,\notag\\
U^{\prime}-\frac{1}{2\sqrt{2}}e^{-2U-\frac{3}{2}\phi}\,\left(e^{2U}(1+3 e^{2\phi})-\sqrt{9+6\sqrt{3}}(e^{\phi}-e^{3\phi})\right)=0\,,\notag\\
\phi^{\prime}+\frac{1}{3\sqrt{2}}e^{-2U-\frac{3}{2}\phi}\,\left(-3\,e^{2U}(1-e^{2\phi})+\sqrt{9+6\sqrt{3}}(e^{\phi}+3e^{3\phi})\right)=0\,.
\end{align}

The expansion close to the $AdS_2\times\mathbb{R}^2$ fixed point in the far IR is 
\begin{align}\label{irexp}
W&=w_{0}+L^{-1}\rho-2\frac{3+\sqrt{3}}{3+2\sqrt{3}}c_{IR}\,e^{L^{-1}\rho}+\cdots\,,\notag\\
U&=c_{IR}\,e^{L^{-1}\rho}+\cdots\,,\notag\\
\phi&=-\frac{1}{4}\,\ln\left[3\left(7+4\sqrt{3} \right) \right]+\frac{2}{2+\sqrt{3}}c_{IR}\,e^{L^{-1}\rho}+\cdots\,,
\end{align}
with $c_{IR}$ and $w_{0}$ being constants of integration.
Setting the magnetic charges $q^i=0$ in the flow equations \eqref{eq:SUSYflow}, we recover the $AdS_{4}$ solution
\begin{equation}
W=U=R^{-1}\,\rho,\quad \phi=0,\quad R^{-1}=\sqrt{2}\,.
\end{equation}
Turning on a non-zero $q^i$ triggers the following asymptotic expansion to the equations \eqref{eq:SU3flow} given by
\begin{align}\label{uvexp}
W&=R^{-1}\,\rho-\frac{3}{16}c_{UV}^{2}\,e^{-2R^{-1}\rho}+\cdots\,,\notag\\
U&=R^{-1}\,\rho-\frac{3}{16}c_{UV}^{2}\,e^{-2R^{-1}\rho}+\cdots\,,\notag\\
\phi&=c_{UV}\,e^{-R^{-1}\,\rho}+\left(2\sqrt{1+\frac{2}{\sqrt{3}}}-\frac{c_{UV}^{2}}{2} \right)\,e^{-2R^{-1}\,\rho}\,,
\end{align}
where $c_{UV}$ is a constant of integration. This expansion corresponds to the operator dual to $\phi$ having a deformation as well as a VEV, and we see that both the deformation and the VEV 
are fixed by $c_{UV}$. 

\begin{figure}
\centering
\includegraphics[width=7cm]{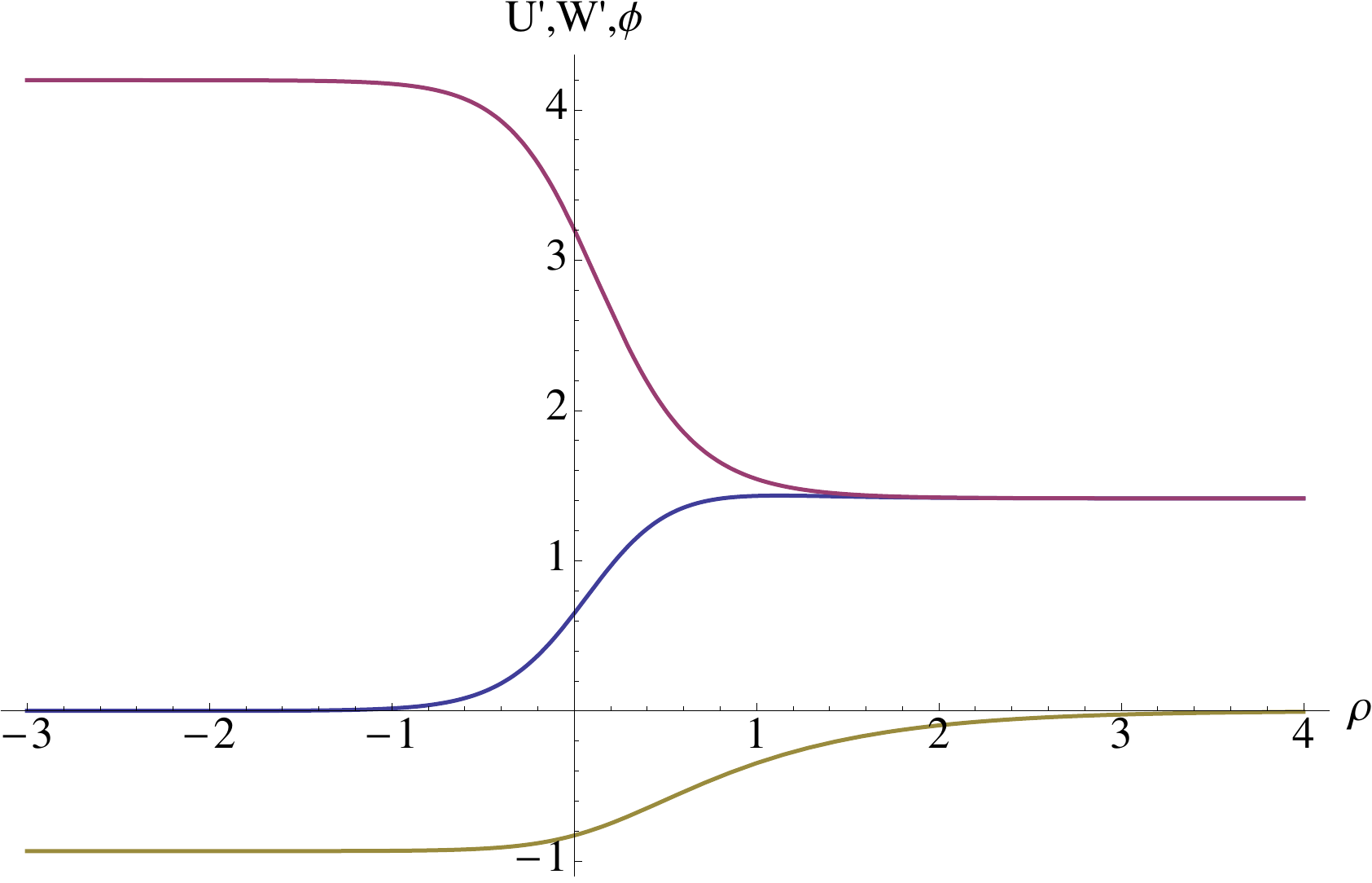}
\caption{We have plotted $U^{\prime}$ (blue) $W^{\prime}$ (purple) and $\phi$ (green) as functions of $\rho$ for the $U(1)^{2}\times SU(3)$ invariant  supersymmetric domain wall solution interpolating between $AdS_{4}$ and $AdS_{2}\times\mathbb{R}^2$.}\label{fig:3}
\end{figure}

Using a shooting method we find that
there is a solution to \eqref{eq:SUSYflow} with boundary conditions \eqref{irexp} and 
\eqref{uvexp} with
\begin{equation}
w_{0}=-0.47\ldots,\quad c_{IR}=0.26\ldots,\quad\,c_{UV}=-1.71\ldots\,,
\end{equation}
as we have indicated in figure \ref{fig:3}.
This is the supersymmetric domain wall solution.
This solution can be uplifted on $S^7$, or an orbifold thereof, to $D=11$ supergravity using the formulae in  \cite{Cvetic:1999xp}. 
The uplifted solutions then describe the dual $d=3$ SCFTs deformed by the presence of the magnetic field and also by the operators dual to $\phi$. In particular,
the supersymmetric $AdS_2\times\mathbb{R}^2$ solutions describes the IR ground state at zero temperature.

\section{Instabilities of magnetic $AdS_2\times\mathbb{R}^2$ solutions}
The instabilities for the magnetic $AdS_2\times\mathbb{R}^2$ solutions that we constructed in section \ref{mag11}
are very similar to those that we have discussed for the
$AdS_3\times\mathbb{R}^2$ solutions in section \ref{section3}. In this section, we will just present some illustrative calculations.

\subsection{Spatially modulated instabilities of neutral scalars}\label{spatmodn11}
We first investigate the possibility of spatially modulated instabilities of the neutral scalars $\phi_a$ for the
magnetic $AdS_2\times\mathbb{R}^2$ solutions given in \eqref{eq:ads2sol}. 
For simplicity we just analyse the one-dimensional subspace of $SU(3)\times U(1)^2$ 
solutions given in \eqref{su3invt}. In particular, we focus on the perturbation with
\begin{align}
\delta\phi_{1}&=0,\qquad \qquad \quad
\delta\phi_{2}=-\delta\phi_{3}=\phi(t,\rho)\,\cos(kx_{1})\,,\notag\\
\delta A^{1}&=\delta A^{2}=0,\qquad 
\delta A^{3}=-\delta A^{4}=a(t,\rho)\,\sin(kx_{1})\,dx_{2}\,.
\end{align}
Defining the vector $\mathbf{v}=\left(\phi,a \right)$, the equations of motion for the Lagrangian \eqref{eq:4dLag} imply,
at linear order,
\begin{align}
&\left(\Box_{AdS_{2}} -L^{2}M^{2}\right)\mathbf{v}=0\,,
\end{align}
where the Laplacian is with respect to a unit radius $AdS_{2}$ and the mass matrix is
\begin{align}
M^{2}=\left(
   \begin{matrix} 
     k^{2} +2X^2+{6}{X^{-2}}&{8  q^2}{X^{-2}}\,k \\
      q^2\,k & k^{2} \\
   \end{matrix}
   \right)\, .
\end{align}
The matrix $M^{2}$ has eigenvalues
\begin{equation}
m_{\pm}^{2}=\frac{1}{X^2}\,\left[3+k^{2}\bar{X}^{2}+\bar{X}^{4}\pm\sqrt{8k^{2}\bar{X}^{2}\left(2+\bar{X}^{4} \right)+\left(3+\bar{X}^{4} \right)^{2}}\right]\,.
\end{equation}
The branch $m_{-}^{2}$ develops a minimum at
\begin{equation}
k_{min}^{2}=\frac{1}{8\bar{X}^{2}}\frac{55+58\bar{X}^{4}+15\bar{X}^{8}}{2+\bar{X}^{4}}\,,
\end{equation}
with the corresponding minimum satisfying
\begin{equation}
L^2m_{min}^{2}=-\frac{1}{48}\,\frac{\left(5+3\bar{X}^{4}\right)^{2}}{2+3\bar{X}^{4}+\bar{X}^{8}}\,.
\end{equation}
For $\bar{X}<\left(-1+\frac{2}{\sqrt{3}}\right)^{1/4}$ the mass minimum violates the $AdS_{2}$ BF bound of $-1/4$ making the solution unstable.

It is worth noting that for the supersymmetric solution with $\bar{X}=\left(-1+\frac{2}{\sqrt{3}}\right)^{1/4}$, the static mode given by
\begin{align}
\phi(\rho)=c_{1}e^{-\frac{\rho}{2L}}\,\cos(\left|k_{min}\right|x_{1}),\qquad
a(r)=c_{2}e^{-\frac{\rho}{2L}}\sin(\left|k_{min}\right|x_{1})\,,
\end{align}
with ${c_{1}^{2}}/{c_{2}^{2}}=8\,\sqrt{-9+6\,\sqrt{3}}$,
which saturates the BF bound, preserves the supersymmetries of the background. The sign choice of $c_{1}$ depends on the choice of $\alpha$ in the projector \eqref{eq:projcond}.

\subsection{Instabilities of charged scalars}
Recall that $SO(8)$ gauged supergravity has 70 scalar fields, parametrising the coset $E_{7(7)}/SU(8)$ and transforming
as two ${\bf 35}$ irreps of $SO(8)$. The scalars in one of these ${\bf 35}$ irreps can be described by
a unimodular $8\times 8$ matrix $T$. The three neutral scalars we have been considering lie in this
irrep. We next investigate possible instabilities of four charged fields lying in this irrep
using the consistent truncation of $SO(8)$ gauge supergravity discussed in \cite{Chong:2004ce}.
We should recall that $A^{there}=2\sqrt{2} A^{here}$ and set their $g=1/\sqrt{2}$. We then follow the earlier analysis in section
\ref{sec1p4} and for the scalar $\varphi_1$ (say) we find
Landau levels with corresponding $AdS_2$ mass given by 
\begin{equation}
m^{2}=-2L^{2}\,\left(X_{1}X_{2}+X_{1}X_{3}+{X_1}{X_4}-X_{1}^{2}-(2n+1)|q_1| \right)\,.
\end{equation}
These modes violate the BF bound for a large parameter space of solutions. 
Indeed, for the lowest level, $n=0$, the unstable regions are, roughly, the obvious
generalisation of figure \ref{fig:AdS3}(b) to figure \ref{3d}. In particular, these modes now
intersect the locus of supersymmetric solutions in a one-dimensional sub-locus.

Finally, we note that there will be additional instabilities for the other scalars in the $8\times 8$ matrix $T$
and these will mix with the gauge fields
and the analysis will mirror the analysis that we carried out in section \eqref{mixedcharged}.

\section{Electric Solutions}
In this section we construct new electric $AdS_2\times\mathbb{R}^2$ and $AdS_2\times\mathbb{R}^3$ solutions.
\subsection{Electric $AdS_2\times\mathbb{R}^2$ solutions}

The equations of motion of the $U(1)^{4}$ truncation of $D=4$ $SO(8)$ gauged supergravity \eqref{eq:4dLag}
are invariant under the electric-magnetic duality transformation
\begin{align}\label{dt}
F^i\to X_i^{-2}*F^i,\qquad \phi_a\to -\phi_a\,,
\end{align}
with the metric unchanged. We can use this symmetry to immediately obtain electric analogues of the magnetic solutions that we presented in section \ref{mag11}.

Starting with \eqref{eq:AdS2ansatz}, \eqref{eq:ads2sol} we obtain electric $AdS_2\times\mathbb{R}^2$ solutions, which we can write as
\begin{align}\label{eq:AdS2ansatze}
ds^{2}&=L^{2}\,ds^{2}\left(AdS_{2}\right)+dx_{1}^{2}+dx_{2}^{2}\,,\notag\\
F^{i}&=\tfrac{1}{2}Q^{i}L^2Vol(AdS_2)\,,\nn
\phi_1&=f_1,\quad \phi_2=f_2,\quad\phi_3=f_3\,,
\end{align}
where 
\begin{align}
\left(Q^{i}\right)^{2}=\bar{X}_{i}^{3}\sum_{j\neq i} \bar{X}_{j},\qquad
L^{-2}=-2\, V\left(\bar{X}_{i} \right)\,,
\end{align}
and, as before,
\begin{align}\label{exse}
\bar X_{1}&=e^{\frac{1}{2}\left(-f_{1}-f_{2}-f_{3} \right)},\quad \bar X_{2}=e^{\frac{1}{2}\left(-f_{1}+f_{2}+f_{3} \right)},\quad \bar X_{3}=e^{\frac{1}{2}\left(f_{1}-f_{2}+f_{3} \right)},\quad \bar X_{4}=e^{\frac{1}{2}\left(f_{1}+f_{2}-f_{3} \right)}\,.
\end{align}
None of the solutions preserve the supersymmetry transformations given in \eqref{gensusvar2}.
For the special case of the electric $AdS_2\times\mathbb{R}^2$ solution of minimal gauged supergravity see
\cite{Denef:2009tp} for a discussion on instabilities. An analysis for other solutions will be carried 
out elsewhere.

Starting with the supersymmetric domain wall solution that we presented in section \ref{sdw4}, we can use the duality transformation \eqref{dt}
to immediately obtain an electric domain wall solution
that interpolates between $AdS_4$ in the UV and an electric $AdS_2\times\mathbb{R}^2$ solution in the IR. 
Note that despite the domain wall not preserving supersymmetry it solves first order flow
equations.

\subsection{Electric $AdS_2\times\mathbb{R}^3$ solutions}
We now consider electric $AdS_2\times\mathbb{R}^3$ solutions of $D=5$ $SO(6)$ gauged supergravity.
By direct construction we find
\begin{align}\label{eq:AdS2_fame}
ds^{2}&=L^{2}\,ds^{2}\left(AdS_{2}\right)+dx_{1}^{2}+dx_{2}^{2}+dx_3^2\,,\notag\\
F^{i}&=2Q^{i}L^2 Vol(AdS_2),\quad\phi_{1}=f_{1},\quad\phi_{2}=f_{2}\,,
\end{align}
where $f_a$ are constants, 
\begin{align}\label{eq:AdS2_constantse}
(Q^{i})^2=\bar{X}_{i}^{3}\,\sum_{j\neq i} \bar{X}_{j},\qquad
L^{-2}=-V(\bar X_i)\,,
\end{align}
and $\bar X_i$ are the on-shell values
\begin{align}
\bar X_{1}&=e^{-\frac{1}{\sqrt{6}}f_{1}-\frac{1}{\sqrt{2}}f_{2}},\quad \bar X_{2}=e^{-\frac{1}{\sqrt{6}}f_{1}+\frac{1}{\sqrt{2}}f_{2}},
\quad \bar X_{3}=e^{\frac{2}{\sqrt{6}}f_{1}}\,.
\end{align}
When $f_2=0$, for example, these are solutions to Romans' theory, and actually were already presented in 
\cite{Romans:1985ps} and further discussed in appendix B of \cite{Donos:2011ff}.
When $f_1=f_2=0$ we obtain the standard $AdS_2\times\mathbb{R}^3$ solution of minimal gauged supergravity which is the near horizon
limit of the usual electrically charged AdS-RN black brane solution. 
Note that the solutions do not preserve the supersymmetry \eqref{susyd51}; within Romans' theory this was shown in \cite{Romans:1985ps}.

It was shown in \cite{Donos:2011ff} that the electrically charged $AdS_2\times\mathbb{R}^3$ solutions in Romans' theory all suffer from instabilities corresponding to helical $p$-wave superconductors. A more detailed stability analysis of all solutions will be carried out elsewhere.

\section*{Acknowledgements}
We would like to thank Nikolay Bobev and Joe Polchinski
for helpful discussions. AD is supported by an EPSRC Postdoctoral Fellowship.
JPG is supported by a Royal Society Wolfson Award. CP is supported by an I.K.Y. Scholarship.

\appendix

\section{Charged mode analysis for the truncation \cite{Bobev:2010de}}\label{22scalars}
A consistent truncation that supplements the $U(1)^3\subset SO(6)$ truncation \eqref{eq:SO6_bh} 
with four complex scalars $\zeta_m$, $m=1,2,3,4$
was given in \cite{Bobev:2010de}. After writing $\zeta_m=\tanh(\gamma_m)e^{i\theta_m}$, the Lagrangian is given
in eq. (2.7) of  \cite{Bobev:2010de} and, to make contact with our notation,  one should set $\varphi^{there}_m=\gamma^{here}_m$,
$g^{there}=2$ and also identify the gauge fields via
$A^{i(there)}=A^{i(here)}/2$. It is straightforward to see that after expanding the equations of motion
around the $AdS_3\times \mathbb{R}^2$ solutions \eqref{eq:AdS2_fam}, the fields $\theta_m$ are
all massless. The analysis for the charged modes $\gamma_m$ is very
similar to that in section \ref{sec1p4}. For example, for $m=1$ we find that the lowest mass mode is obtained by writing
\begin{equation}
\gamma_{1}=e^{-\frac{|q_1+q_2-q_3|}{4}\,\left(x_{1}^{2}+x_{2}^{2}\right)}\,\sigma_{1}(t,r,z)\,,
\end{equation}
 giving
 \begin{equation}
(\Box_{AdS_{3}}-L^{2}m_{\sigma_{1}}^{2})\sigma_{1}=0\,,
\end{equation}
with
\begin{equation}
m_{\sigma_{1}}^{2}=|q_1+q_2-q_3|+\sum_i \bar X_i^2-2\sum_i \bar X_i^{-1}\,.
\end{equation}
Over the moduli space of $AdS_3\times \mathbb{R}^2$ slutions we find that the minimum value is $L^2m_{\sigma_{1}}^{2}\approx -0.704$ and does not violate the BF bound.

\section{The mixed charged modes}\label{appendixcharged}
Here we provide some details of the calculations we carried out for section \ref{mixedcharged}. We consider the perturbation $t,a$ 
about the background $AdS_3\times\mathbb{R}^2$ solution \eqref{eq:AdS2_fam},\eqref{eq:AdS2_constants}  
defined by
\begin{align}
 T=\tilde{T}+t,\qquad A=\tilde A +a\,,
\end{align}
with
\begin{align}\label{bground}
\tilde{T}_{I\bar{I}}=X_{I},\qquad 
\tilde{A}_{I\bar{I}}=q_I\left(\bar{z}\,dz-z\,d\bar{z} \right)\,.
\end{align}
It is is useful to note that at leading order in the perturbation we have $T^{-1}=\tilde{T}^{-1}-\tilde{T}^{-1}t\tilde{T}^{-1}$. Furthermore, the
linearised expression of the field strengths and some covariant derivatives are given by
\begin{align}\label{longlist}
\delta(F_{IJ})&=da_{IJ}+\,\left(\tilde{A}_{I\bar{I}}+\tilde{A}_{J\bar{J}} \right)\wedge a_{IJ}\,,\nn
\delta(F_{I\bar{J}})&=da_{I\bar{J}}+\,\left(\tilde{A}_{I\bar{I}}-\tilde{A}_{J\bar{J}} \right)\wedge a_{I\bar{J}}\,,\nn
\delta(DT_{IJ})&=dt_{IJ}+\,\left(\tilde{A}_{I\bar{I}}+\tilde{A}_{J\bar{J}} \right)\,t_{IJ}+g\,\left(X_{J}-X_{I} \right)\,a_{IJ}\,,\nn
\delta(DT_{I\bar{J}})&=dt_{I\bar{J}}+\left(\tilde{A}_{I\bar{I}}-\tilde{A}_{J\bar{J}} \right)\,t_{I\bar{J}}+g\,\left(X_{J}-X_{I} \right)\,a_{I\bar{J}}\,,\nn
\delta(D\ast F_{IJ})&=d\ast \delta(F_{IJ})+\,\left(\tilde{A}_{I\bar{I}}+\tilde{A}_{J\bar{J}}\right)\wedge\ast \delta(F_{IJ})+\,\ast\left(\tilde{F}_{I\bar{I}}+\tilde{F}_{J\bar{J}} \right)\wedge a_{IJ}\,,
\nn
\delta(D\ast DT_{IJ})&=d\ast \delta(Dt_{IJ})+\left(\tilde{A}_{I\bar{I}}+\tilde{A}_{J\bar{J}}\right)\wedge\ast \delta(Dt_{IJ})\,.
\end{align}

We will only provide details concerning the  $IJ$ components of the equations of motion \eqref{eq:vectoreq}, with $I\ne J$.
The case of $I\bar J$. is very similar.
At linearised order we have
\begin{align}
& \delta(D\ast F_{IJ})+\left(X^{-1}_{I}\,\ast \tilde{F}_{I\bar{I}}-X^{-1}_{J}\,\ast \tilde{F}_{J\bar{J}}\right)\wedge \delta(Dt_{IJ})
=-\left(X_{J}-X_{I}\right)\,\ast \delta(Dt_{IJ})\,,\label{eq:pertveceom}\\
& -\delta(D\ast Dt_{IJ})=4q_I^{-1}q_J^{-1}\,t_{IJ}+\ast\left(X_{J}^{-1}\tilde{F}_{J\bar{J}}- X_{I}^{-1}\tilde{F}_{I\bar{I}}\right)\wedge \delta(F_{IJ})\,.\label{eq:pertsceom}
\end{align}
For the gauge fields we take
\begin{equation}
a_{IJ}=a^{1}_{IJ}\,dz+a^{2}_{IJ}\,d\bar{z}\,,
\end{equation}
and, after defining 
\begin{align}
\omega_{IJ}=q_I+q_J\,,
\end{align} we find
\begin{align}
\delta(F_{IJ})&=da_{IJ}^{1}\wedge dz+da_{IJ}^{2}\wedge d\bar{z}-\left(\partial_{\bar{z}}-z\omega_{IJ}\right)a_{IJ}^{1}\,dz\wedge d\bar{z}+\left(\partial_{z}+\bar{z}\omega_{IJ}\right)a_{IJ}^{2}\,dz\wedge d\bar{z}\,,\notag\\
\delta(Dt_{IJ})&=dt_{IJ}+\left(\partial_{z}+\bar{z}\omega_{IJ}\right)t_{IJ}\,dz+\left(\partial_{\bar{z}}-z\omega_{IJ}\right)t_{IJ}\,d\bar{z}+\left(X_{J}-X_{I} \right)\left(a_{IJ}^{1}\,dz+a_{IJ}^{2}\,d\bar{z} \right)\,,
\end{align}
where $d$ is the exterior derivative on $AdS_{3}$. We can obtain analogous expressions for 
$\delta(D\ast F_{IJ})$ and $\delta(D\ast Dt_{IJ})$ using \ref{longlist}, which we then substitute into 
the equations of motion \eqref{eq:pertveceom} and \eqref{eq:pertsceom}.

From \eqref{eq:pertveceom} we are led to impose the constraint
\begin{equation}\label{eq:constrain}
\left(\partial_{\bar{z}}-\omega_{IJ}z \right)a_{IJ}^{1}+\left(\partial_{z}+\omega_{IJ}\bar{z} \right)a_{IJ}^{2}=-\left(X_{J}-X_{I} \right)\,t_{IJ}\,.
\end{equation}
We also obtain 
\begin{align}\label{eq:a1eq}
&-L^{-2}\Box_{AdS_3} a_{IJ}^{1}+2\,\left(-\partial_{z}\partial_{\bar{z}}+\omega_{IJ}^{2}z\bar{z}+2\omega_{IJ}+\omega_{IJ}\,\left(z\partial_{z}-\bar{z}\partial_{\bar{z}} \right)\right)a_{IJ}^{1}\notag\\
&+2\left(q_I^{-1} -q_J^{-1} \right)\,\left[\left(\partial_{z}+\omega_{IJ}\bar{z} \right)t_{IJ}+\left(X_{J}-X_{I}\right)a_{IJ}^{1} \right]=-\left(X_{J}-X_{I} \right)^{2}\,a_{IJ}^{1}\,,
\end{align}
and
\begin{align}\label{eq:a2eq}
&L^{-2}\Box_{AdS_3}a_{IJ}^{2}+2\,\left(\partial_{z}\partial_{\bar{z}}-\omega_{IJ}^{2}z\bar{z}+2\omega_{IJ}-\omega_{IJ}\,\left(z\partial_{z}-\bar{z}\partial_{\bar{z}} \right)\right)a_{IJ}^{2}\notag\\
&+2\left(q_I^{-1} -q_J^{-1}\right)\,\left[\left(\partial_{\bar{z}}-\omega_{IJ}z \right)t_{IJ}+\left(X_{J}-X_{I}\right)a_{IJ}^{2} \right]=\left(X_{J}-X_{I} \right)^{2}\,a_{IJ}^{2}\,,
\end{align}
where $\Box_{AdS_3}$ is the Laplacian on a unit radius $AdS_{3}$.
Similarly, from \eqref{eq:pertsceom} we obtain
\begin{align}\label{eq:teqng}
&L^{-2}\Box_{AdS_3}t_{IJ}+2\left(\partial_{z}\partial_{\bar{z}}-\omega_{IJ}^{2}z\bar{z}+\omega_{IJ}\,\left(\bar{z}\partial_{\bar{z}}-z\partial_{z} \right) \right)t_{IJ}-\left(X_{J}-X_{I}\right)^{2}t_{IJ}=\nn&-4q_I^{-1}q_J^{-1}t_{IJ}
+2\,\left(q_I^{-1} -q_J^{-1} \right)\,\left[\left(-\partial_{\bar{z}}+\omega_{IJ}z \right)a_{IJ}^{1}+\left(\partial_{z}+\omega_{IJ}\bar{z} \right)a_{IJ}^{2} \right]\,.
\end{align}
We now observe that because of the constraint \eqref{eq:constrain}, the three equations
\eqref{eq:a1eq}-\eqref{eq:teqng} are not independent. Indeed 
acting on equation \eqref{eq:a1eq} by $\left(-\partial_{\bar{z}}+\omega_{IJ}z \right)$, on equation \eqref{eq:a2eq} by $\left(\partial_{z}+\omega_{IJ}\bar{z} \right)$ and adding one can show that equation \eqref{eq:teqng} is satisfied. 

To continue with the analysis, we need to fix the sign of $\omega\equiv \omega_{IJ}$. We first take $\omega> 0$. For this case we can
keep equation \eqref{eq:a2eq} and \eqref{eq:teqng} which we write as
\begin{align}\label{eq:teqn}
& L^{-2}\Box_{AdS_3} t_{IJ}+2\left(\partial_{z}\partial_{\bar{z}}-\omega^{2}z\bar{z}+\omega\,\left(\bar{z}\partial_{\bar{z}}-z\partial_{z} \right) \right)t_{IJ}-\left(X_{J}-X_{I}\right)^{2}t_{IJ}=\nn&-4q_I^{-1}q_J^{-1}t_{IJ}
+2\,\left(q_I^{-1} -q_J^{-1} \right)\,\left[\left(X_{J}-X_{I} \right)t_{IJ}+2\left(\partial_{z}+\omega\bar{z} \right)a_{IJ}^{2} \right] \,.
\end{align}
Next we introduce the ladder operators
\begin{align}\label{ladderops}
&a=\frac{1}{\sqrt{2\omega}}\,\left(\partial_{\bar{z}}+\omega\,z \right),\quad a^{\dag}=\frac{1}{\sqrt{2\omega}}\,\left(-\partial_{z}+\omega\,\bar{z} \right)\,,\nn
&b=\frac{1}{\sqrt{2\omega}}\,\left(\partial_{z}+\omega\,\bar{z}\right),\quad b^{\dag}=\frac{1}{\sqrt{2\omega}}\,\left(-\partial_{\bar{z}}+\omega\,z\right)\,,
\end{align}
which can be checked to satisfy the algebra
\begin{equation}
\left[a,a^{\dag} \right]=1,\quad \left[b,b^{\dag}\right]=1\,,
\end{equation}
and the rest of the commutators being trivial. Note that we have
\begin{align}
-\partial_{z}\partial_{\bar{z}}+\omega^{2}\bar{z}z&=\omega\,\left(a^{\dag}a+b^{\dag}b+1 \right)\,, \nn
\bar{z}\partial_{\bar{z}}-z\partial_{z}&=a^{\dag}a-b^{\dag}b\,.
\end{align}
In terms of these operators, equations \eqref{eq:a2eq} and \eqref{eq:teqn} take the form
\begin{align}\label{eq:pert1}
&L^{-2}\Box_{AdS_3}a_{IJ}^{2}-2\omega\,\left(2b^{\dag}b-1\right)a_{IJ}^{2}\notag\\
&+2\left(q_I^{-1} -q_J^{-1}  \right)\,\left[-\sqrt{2\omega}\,b^{\dag}\,t_{IJ}+\left(X_{J}-X_{I}\right)a_{IJ}^{2} \right]-\left(X_{J}-X_{I} \right)^{2}\,a_{IJ}^{2}=0
\end{align}
and
\begin{align}\label{eq:pert2}
& L^{-2}\Box_{AdS_3} t_{IJ}-2\omega\,\left(2b^{\dag}b+1\right)t_{IJ}-\left(X_{J}-X_{I}\right)^{2}t_{IJ}+4q_I^{-1}q_J^{-1}t_{IJ}\notag\\
&-2\,\left(q_I^{-1} -q_J^{-1}  \right)\,\left[\left(X_{J}-X_{I} \right)t_{IJ}+2\sqrt{2\omega}\,b\,a_{IJ}^{2} \right]=0\,.
\end{align}

To reduce the problem to modes on the $AdS_3$ space we introduce the ground state
\begin{equation}
L_{0,0}=\left(\frac{\omega}{\pi} \right)^{1/2}\,e^{-\omega z\bar{z}}
\end{equation}
and the complete set of functions
\begin{equation}
L_{n,m}\left(z,\bar{z}\right)=\frac{\left(b^{\dag}\right)^{n}}{\sqrt{n!}}\frac{\left(a^{\dag}\right)^{m}}{\sqrt{m!}}\,L_{0,0}\left(z,\bar{z}\right),\quad m,n>0\,.
\end{equation}
We use these to write the expansions
\begin{align}
t_{IJ}=\sum_{n,m}\,f_{IJ}^{n,m}\,L_{n,m}\left(z,\bar{z}\right),\qquad 
a_{IJ}^{2}=\sum_{n,m}\,g_{IJ}^{n,m}\,L_{n,m}\left(z,\bar{z}\right)\,,
\end{align}
with $f$ and $g$ defined on $AdS_{3}$. From equations \eqref{eq:pert1} and \eqref{eq:pert2} we see that the modes $g_{IJ}^{0,m}$ decouple and they have an $AdS_{3}$ mass
\begin{equation}\label{pip}
m_{0,m}^{2}=-2\omega-2\,\left(q_I^{-1} -q_J^{-1} \right)\left(X_{J}-X_{I}\right)+\left(X_{J}-X_{I}\right)^{2}\,.
\end{equation}

For the rest of the modes, we see that $g_{IJ}^{n+1,m}$ mix with $f_{IJ}^{n,m}$ for $n\geq 0$ with mass matrix
\begin{align}\label{eq:mass_matrix2}
M^{2}=\left(
   \begin{matrix} 
     2\omega\left(2n+1\right)-4q_I^{-1}q_J^{-1}+2W_{IJ}V_{JI} +V_{IJ}^{2}& 4\sqrt{2\omega}W_{IJ}(n+1)^{1/2} \\
     2\sqrt{2\omega}W_{IJ}\,\left(n+1\right)^{1/2} & 2\omega\left(2n+1\right)-2W_{IJ}V_{JI}+V_{IJ}^{2} \\
   \end{matrix}
   \right)\, ,
\end{align}
where we set
\begin{align}
W_{IJ}=q_I^{-1} -q_J^{-1},\qquad
V_{IJ}=X_{I}-X_{J}\,.
\end{align}
Note that our results differ from those presented in the first two lines of
eq. (6.7) and eq. (6.12) of \cite{Almuhairi:2011ws} because the mixing of the charged scalars and vectors was not taken into account in that reference\footnote{Our zero modes \eqref{pip} only involve
the gauge fields and the fact that our results differ from eq. (6.12) of \cite{Almuhairi:2011ws} after setting their $n=0$, which also just involve
the gauge-fields, can be traced back to the fact that our equation of motion \eqref{eq:pertveceom} for $t_{IJ}=0$ does not come from the Lagrangian given in eq. (6.11) of
\cite{Almuhairi:2011ws}.}.

When $\omega<0$, in order to get the zero modes, we should keep equations \eqref{eq:a1eq} and \eqref{eq:teqng}. For the ladder
operators we should take $\sqrt{2\omega}\to\sqrt{2|\omega|}$ and $\omega\to-\omega$ in \eqref{ladderops}. A very similar analysis then ensues
and we obtain \eqref{pip} and \eqref{eq:mass_matrix2} after substituting $\omega\to |\omega|$ and also $W_{IJ}\to - W_{IJ}$.

Finally, let us consider $\omega=0$. This occurs along the three lines in figure \ref{fig:susy} with $X_I=X_J$. The equations
\eqref{eq:constrain} - \eqref{eq:teqng} then simplify and we are essentially led back to the mass matrix that we saw for
spatially modulated neutral scalars in \eqref{eq:mass_matrix}.

\section{Construction of SUSY  $AdS_{2}\times\mathbb{R}^2$ solutions}\label{dircon}
Recall \cite{Kim:2006qu} that supersymmetric $AdS_2$ solutions of $D=11$ supergravity with purely
electric four-form flux, generically dual to CFTs with two (Poincar\'e) supersymmetries, can be obtained from an eight dimensional K\"ahler metric, $ds^2_8$,
whose Ricci tensor satisfies
\begin{equation}\label{masteq}
\Box_{8}R-\frac{1}{2}R^{2}+{R}_{ij}{R}^{ij}=0\,.
\end{equation}
The $D=11$ metric has the form
\begin{align}\label{metan}
ds^2=e^{2A}\left[ds^2(AdS_2)+e^{-3A}ds^2_8+(dz+P)^2\right]\,,
\end{align} 
where $dP={\cal R}$, where $\cal R$ is the Ricci-form, and $e^{-3A}=\tfrac{1}{2}R$.

Following \cite{Gauntlett:2007sm} we start with an ansatz for an eight dimensional K\"ahler metric given by 
\begin{align}
ds_{8}^{2}&=\frac{dy^{2}}{U}+y^{2}U\,\left(D\phi+A\right)^{2}+y^{2}\,ds^{2}({\mathbb{CP}^{2}})+\left(ay^{2}+b\right)\,ds^{2}({\mathbb{R}^{2}})\,,
\end{align}
with  $dD\phi=2\,J_{\mathbb{CP}^{2}}$, where $J_{\mathbb{CP}^{2}}$ is the K\"ahler form on $\mathbb{CP}^{2}$, and
$dA=2a\,J_{\mathbb{R}^{2}}$, where $J_{\mathbb{R}^{2}}=Vol(\mathbb{R}^2)$ and $a$ is a constant, and $U=U(y)$. 
For the corresponding holomorphic 2-form and 1-form on $\mathbb{CP}^{2}$ and $\mathbb{R}^{2}$ we have
\begin{align}
d\Omega_{\mathbb{CP}^{2}}&=i\,P_{\mathbb{CP}^{2}}\wedge\Omega_{\mathbb{CP}^{2}},\quad dP_{\mathbb{CP}^{2}}=2l\,J_{\mathbb{CP}^{2}}\,,\nn
d\Omega_{\mathbb{R}^{2}}&=0\,,
\end{align}
where $l$ is another (positive) constant.
The K\"ahler-form and holomorphic 4-form for the eight dimensional space can now be written
\begin{align}
J&=y\,dy\wedge\left(D\phi+A\right)+y^{2}\,J_{\mathbb{CP}^{2}}+\left(ay^{2}+b\right)\,J_{\mathbb{R}^{2}}\,,\nn
\Omega_{4}&=e^{i l\phi}y^2\sqrt{ay^2+b}\left[\frac{dy}{\sqrt{U}}+i y\sqrt{U}\,\left(D\phi+A\right) \right]\wedge\Omega_{\mathbb{CP}^{2}}\wedge\Omega_{\mathbb{R}^{2}}\,.
\end{align}
We can easily show that
\begin{align}
d\Omega_{4}&=i\,P\wedge \Omega_{4},\qquad
P=l\,D\phi-g\,\left(D\phi+A\right)\,,
\end{align}
where
\begin{align}
g&=3U+\frac{ay^2U}{ay^2+b}+\frac{yU'}{2}\,.
\end{align}
The Ricci form for the eight dimensional space is given by
\begin{equation}
\mathcal{R}=dP=2\left(l-g\right)\,J_{\mathbb{CP}^{2}}-2ag\,J_{\mathbb{R}^{2}}-g'\,dy\wedge\left(D\phi+A\right)\,.
\end{equation}

In order to get an $AdS_2\times \mathbb{R}^2$ factor in \eqref{metan}
we now require that the Ricci scalar of the eight dimensional space satisfies
\begin{equation}
R=\frac{W}{ay^{2}+b}\,,
\end{equation}
for some constant $W$. The resulting second order equation for $U$ gives the solution
\begin{equation}
U=\frac{1}{48}\,\frac{1}{ay^{2}+b}\,\left( 16bl+8aly^{2}-Wy^{2}+\frac{c_{1}}{y^{6}}+\frac{c_{2}}{y^{4}}\right)\,,
\end{equation}
where $c_{i}$ are two constants of integration. One can check that in order to solve \eqref{masteq} we need to set $c_1=c_2=0$ and we then find two solutions
\begin{align}
W=&-4\,\left(1\mp\sqrt{3}\right)\,al\quad \Rightarrow \quad  U=\frac{l}{12}\frac{4b+\left(3\mp\sqrt{3}\right)ay^{2}}{b+ay^{2}}\,.\label{eq:sol1}
\end{align}
Let us now continue with the solution with the lower sign. We take $a<0$ and change coordinates via
$y=\frac{2}{\sqrt{3+\sqrt{3}}}\sqrt{-\frac{b}{a}}\,\sin\xi$ and we will take $0<\xi<\pi/2$. We find
\begin{align}
U&=l\left(1+\frac{1}{\sqrt{3}}\right)\,\frac{\cos^{2}\xi}{1+\sqrt{3}+2\cos2\xi}\,,\nn
R_{8}=&-8\frac{al\,\left(3+2\sqrt{3}\right)}{b}\,\frac{1}{1+\sqrt{3}+2\,\cos2\xi}\,,\nn
g=&l\frac{3+\sqrt{3}}{3}\,\frac{1+2\cos 2\xi}{1+\sqrt{3}+2\cos 2\xi}\,.
\end{align}
We can now assemble the $D=11$ metric using \eqref{metan} and, after setting $l=3$, find
\begin{align}
&ds^2=e^{2A}L^{-2}\bigg\{
L^2ds^2(AdS_2)+\frac{L^2W}{2}ds^2(\mathbb{R}^2)\nn
&+2\bar X^2\bigg(d\xi^2+\frac{\sin^2\xi}{\bar X^3\Delta}[ds^2(CP^2)+(D\tilde\phi+\frac{1+\sqrt 3}{4}A)^2]+\frac{\bar X\cos^2\xi}{\Delta}\frac{1}{8\sqrt{3} \bar X^4}(dz-3A)^2\bigg)
\bigg\}
\end{align}
where $\bar X$ is as in \eqref{eq:SU3fp} and
\begin{align}
\Delta=\frac{1}{\bar X^3}(\cos^2\xi+\bar X^4\sin^2\xi)\,.
\end{align}
One can now check that if we set 
\begin{align}-\frac{a}{b}=\frac{4}{3^{1/4}(1+\sqrt 3)}
\end{align}
and rescale $(L^2 W/2)ds^2(\mathbb{R}^2)\to ds^2(\mathbb{R}^2)$ then we precisely obtain the 
uplift of the solution \eqref{eq:SU3fp} using the formulae in \cite{Cvetic:1999xp} (setting $g^2=1/2$ in eq. (3.8) of \cite{Cvetic:1999xp}
and identifying $(F^{i})^{there}=2\sqrt{2}(F^{i})^{here}$).
\bibliographystyle{utphys}
\bibliography{magnstripes}{}

\providecommand{\href}[2]{#2}\begingroup\raggedright\begin{thebibliography}{10}

\bibitem{Hartnoll:2007ai}
S.~A. Hartnoll and P.~Kovtun, ``{Hall conductivity from dyonic black holes},''
  \href{http://dx.doi.org/10.1103/PhysRevD.76.066001}{{\em Phys. Rev.}
  {\bfseries D76} (2007) 066001},
\href{http://arxiv.org/abs/0704.1160}{{\ttfamily arXiv:0704.1160 [hep-th]}}.

\bibitem{Hartnoll:2007ih}
S.~A. Hartnoll, P.~K. Kovtun, M.~Muller, and S.~Sachdev, ``{Theory of the
  Nernst effect near quantum phase transitions in condensed matter, and in
  dyonic black holes},''
  \href{http://dx.doi.org/10.1103/PhysRevB.76.144502}{{\em Phys. Rev.}
  {\bfseries B76} (2007) 144502},
\href{http://arxiv.org/abs/0706.3215}{{\ttfamily arXiv:0706.3215
  [cond-mat.str-el]}}.

\bibitem{Hartnoll:2007ip}
S.~A. Hartnoll and C.~P. Herzog, ``{Ohm's Law at strong coupling: S duality and
  the cyclotron resonance},''
  \href{http://dx.doi.org/10.1103/PhysRevD.76.106012}{{\em Phys. Rev.}
  {\bfseries D76} (2007) 106012},
\href{http://arxiv.org/abs/0706.3228}{{\ttfamily arXiv:0706.3228 [hep-th]}}.

\bibitem{Albash:2008eh}
T.~Albash and C.~V. Johnson, ``{A Holographic Superconductor in an External
  Magnetic Field},''
  \href{http://dx.doi.org/10.1088/1126-6708/2008/09/121}{{\em JHEP} {\bfseries
  0809} (2008) 121}, \href{http://arxiv.org/abs/0804.3466}{{\ttfamily
  arXiv:0804.3466 [hep-th]}}.

\bibitem{Gauntlett:2007sm}
J.~P. Gauntlett and O.~Varela, ``{D=5 $SU(2)$xU(1) Gauged Supergravity from
  D=11 Supergravity},''
  \href{http://dx.doi.org/10.1088/1126-6708/2008/02/083}{{\em JHEP} {\bfseries
  02} (2008) 083},
\href{http://arxiv.org/abs/0712.3560}{{\ttfamily arXiv:0712.3560 [hep-th]}}.

\bibitem{Romans:1985ps}
L.~J. Romans, ``{Gauged N=4 supergravities in five-dimensions and their
  magnetovac backgrounds},''
\href{http://dx.doi.org/10.1016/0550-3213(86)90398-6}{{\em Nucl. Phys.}
  {\bfseries B267} (1986) 433}.

\bibitem{Lin:2004nb}
H.~Lin, O.~Lunin, and J.~M. Maldacena, ``{Bubbling AdS space and 1/2 BPS
  geometries},'' \href{http://dx.doi.org/10.1088/1126-6708/2004/10/025}{{\em
  JHEP} {\bfseries 10} (2004) 025},
\href{http://arxiv.org/abs/hep-th/0409174}{{\ttfamily arXiv:hep-th/0409174}}.

\bibitem{Gaiotto:2009gz}
D.~Gaiotto and J.~Maldacena, ``{The gravity duals of N=2 superconformal field
  theories},''
\href{http://arxiv.org/abs/0904.4466}{{\ttfamily arXiv:0904.4466 [hep-th]}}.

\bibitem{D'Hoker:2009bc}
E.~D'Hoker and P.~Kraus, ``{Charged Magnetic Brane Solutions in $AdS_5$ and the
  fate of the third law of thermodynamics},''
  \href{http://dx.doi.org/10.1007/JHEP03(2010)095}{{\em JHEP} {\bfseries 03}
  (2010) 095},
\href{http://arxiv.org/abs/0911.4518}{{\ttfamily arXiv:0911.4518 [hep-th]}}.

\bibitem{D'Hoker:2010rz}
E.~D'Hoker and P.~Kraus, ``{Holographic Metamagnetism, Quantum Criticality, and
  Crossover Behavior},'' \href{http://dx.doi.org/10.1007/JHEP05(2010)083}{{\em
  JHEP} {\bfseries 05} (2010) 083},
\href{http://arxiv.org/abs/1003.1302}{{\ttfamily arXiv:1003.1302 [hep-th]}}.

\bibitem{D'Hoker:2010ij}
E.~D'Hoker and P.~Kraus, ``{Magnetic Field Induced Quantum Criticality via new
  Asymptotically $AdS_5$ Solutions},''
  \href{http://dx.doi.org/10.1088/0264-9381/27/21/215022}{{\em Class. Quant.
  Grav.} {\bfseries 27} (2010) 215022},
\href{http://arxiv.org/abs/1006.2573}{{\ttfamily arXiv:1006.2573 [hep-th]}}.

\bibitem{D'Hoker:2009mm}
E.~D'Hoker and P.~Kraus, ``{Magnetic Brane Solutions in AdS},''
  \href{http://dx.doi.org/10.1088/1126-6708/2009/10/088}{{\em JHEP} {\bfseries
  10} (2009) 088},
\href{http://arxiv.org/abs/0908.3875}{{\ttfamily arXiv:0908.3875 [hep-th]}}.

\bibitem{Gauntlett:2005ww}
J.~P. Gauntlett, D.~Martelli, J.~Sparks, and D.~Waldram, ``{Supersymmetric
  AdS(5) solutions of type IIB supergravity},''
  \href{http://dx.doi.org/10.1088/0264-9381/23/14/009}{{\em Class. Quant.
  Grav.} {\bfseries 23} (2006) 4693--4718},
\href{http://arxiv.org/abs/hep-th/0510125}{{\ttfamily arXiv:hep-th/0510125}}.

\bibitem{Gauntlett:2004zh}
J.~P. Gauntlett, D.~Martelli, J.~Sparks, and D.~Waldram, ``{Supersymmetric
  AdS(5) solutions of M-theory},''
  \href{http://dx.doi.org/10.1088/0264-9381/21/18/005}{{\em Class. Quant.
  Grav.} {\bfseries 21} (2004) 4335--4366},
\href{http://arxiv.org/abs/hep-th/0402153}{{\ttfamily arXiv:hep-th/0402153}}.

\bibitem{Gauntlett:2006ai}
J.~P. Gauntlett, E.~O~Colgain, and O.~Varela, ``{Properties of some conformal
  field theories with M-theory duals},''
  \href{http://dx.doi.org/10.1088/1126-6708/2007/02/049}{{\em JHEP} {\bfseries
  0702} (2007) 049}, \href{http://arxiv.org/abs/hep-th/0611219}{{\ttfamily
  arXiv:hep-th/0611219 [hep-th]}}.

\bibitem{Gauntlett:2007ma}
J.~P. Gauntlett and O.~Varela, ``{Consistent Kaluza-Klein Reductions for
  General Supersymmetric AdS Solutions},''
  \href{http://dx.doi.org/10.1103/PhysRevD.76.126007}{{\em Phys. Rev.}
  {\bfseries D76} (2007) 126007},
\href{http://arxiv.org/abs/0707.2315}{{\ttfamily arXiv:0707.2315 [hep-th]}}.

\bibitem{Almuhairi:2010rb}
A.~Almuhairi, ``{AdS$_3$ and AdS$_2$ Magnetic Brane Solutions},''
\href{http://arxiv.org/abs/1011.1266}{{\ttfamily arXiv:1011.1266 [hep-th]}}.

\bibitem{Almuhairi:2011ws}
A.~Almuhairi and J.~Polchinski, ``{Magnetic $AdS \times R^2$: Supersymmetry and
  stability},''
\href{http://arxiv.org/abs/1108.1213}{{\ttfamily arXiv:1108.1213 [hep-th]}}.

\bibitem{Nakamura:2009tf}
S.~Nakamura, H.~Ooguri, and C.-S. Park, ``{Gravity Dual of Spatially Modulated
  Phase},'' \href{http://dx.doi.org/10.1103/PhysRevD.81.044018}{{\em Phys.
  Rev.} {\bfseries D81} (2010) 044018},
\href{http://arxiv.org/abs/0911.0679}{{\ttfamily arXiv:0911.0679 [hep-th]}}.

\bibitem{Donos:2011bh}
A.~Donos and J.~P. Gauntlett, ``{Holographic striped phases},''
  \href{http://dx.doi.org/10.1007/JHEP08(2011)140}{{\em JHEP} {\bfseries 08}
  (2011) 140},
\href{http://arxiv.org/abs/1106.2004}{{\ttfamily arXiv:1106.2004 [hep-th]}}.

\bibitem{Bergman:2011rf}
O.~Bergman, N.~Jokela, G.~Lifschytz, and M.~Lippert, ``{Striped instability of
  a holographic Fermi-like liquid},''
\href{http://arxiv.org/abs/1106.3883}{{\ttfamily arXiv:1106.3883 [hep-th]}}.

\bibitem{Donos:2011ff}
A.~Donos and J.~P. Gauntlett, ``{Holographic helical superconductors},''
\href{http://arxiv.org/abs/1109.3866}{{\ttfamily arXiv:1109.3866 [hep-th]}}.

\bibitem{Ammon:2011je}
M.~Ammon, J.~Erdmenger, P.~Kerner, and M.~Strydom, ``{Black Hole Instability
  Induced by a Magnetic Field},''
\href{http://arxiv.org/abs/1106.4551}{{\ttfamily arXiv:1106.4551 [hep-th]}}.

\bibitem{Chernodub:2010qx}
M.~N. Chernodub, ``{Superconductivity of QCD vacuum in strong magnetic
  field},'' \href{http://dx.doi.org/10.1103/PhysRevD.82.085011}{{\em Phys.
  Rev.} {\bfseries D82} (2010) 085011},
\href{http://arxiv.org/abs/1008.1055}{{\ttfamily arXiv:1008.1055 [hep-ph]}}.

\bibitem{Nielsen:1978rm}
N.~K. Nielsen and P.~Olesen, ``{An Unstable Yang-Mills Field Mode},''
\href{http://dx.doi.org/10.1016/0550-3213(78)90377-2}{{\em Nucl. Phys.}
  {\bfseries B144} (1978) 376}.

\bibitem{Ambjorn:1979xi}
J.~Ambjorn and P.~Olesen, ``{On the Formation of a Random Color Magnetic
  Quantum Liquid in QCD},''
\href{http://dx.doi.org/10.1016/0550-3213(80)90476-9}{{\em Nucl. Phys.}
  {\bfseries B170} (1980) 60}.

\bibitem{Rasolt:1992zz}
M.~Rasolt and Z.~Tesanovic, ``{Theoretical aspects of superconductivity in very
  high magnetic fields},''
\href{http://dx.doi.org/10.1103/RevModPhys.64.709}{{\em Rev. Mod. Phys.}
  {\bfseries 64} (1992) 709--754}.

\bibitem{urrhge}
D.~Aoki, T.~D. Matsuda, V.~Taufour, E.~Hassinger, G.~Knebel, and J.~Flouquet,
  ``{Superconductivity Reinforced by Magnetic Field and the Magnetic
  Instability in Uranium Ferromagnets},'' {\em J. Phys. Soc. Jpn.} {\bfseries
  78} (2009) 113709, \href{http://arxiv.org/abs/1012.1987}{{\ttfamily
  arXiv:1012.1987 [cond-mat]}}.

\bibitem{Lee:2008xf}
S.-S. Lee, ``{A Non-Fermi Liquid from a Charged Black Hole: A Critical Fermi
  Ball},'' \href{http://dx.doi.org/10.1103/PhysRevD.79.086006}{{\em Phys.Rev.}
  {\bfseries D79} (2009) 086006},
  \href{http://arxiv.org/abs/0809.3402}{{\ttfamily arXiv:0809.3402 [hep-th]}}.

\bibitem{Liu:2009dm}
H.~Liu, J.~McGreevy, and D.~Vegh, ``{Non-Fermi liquids from holography},''
  \href{http://dx.doi.org/10.1103/PhysRevD.83.065029}{{\em Phys.Rev.}
  {\bfseries D83} (2011) 065029},
  \href{http://arxiv.org/abs/0903.2477}{{\ttfamily arXiv:0903.2477 [hep-th]}}.

\bibitem{Cubrovic:2009ye}
M.~Cubrovic, J.~Zaanen, and K.~Schalm, ``{String Theory, Quantum Phase
  Transitions and the Emergent Fermi-Liquid},''
  \href{http://dx.doi.org/10.1126/science.1174962}{{\em Science} {\bfseries
  325} (2009) 439--444}, \href{http://arxiv.org/abs/0904.1993}{{\ttfamily
  arXiv:0904.1993 [hep-th]}}.

\bibitem{Faulkner:2009wj}
T.~Faulkner, H.~Liu, J.~McGreevy, and D.~Vegh, ``{Emergent quantum criticality,
  Fermi surfaces, and AdS(2)},''
  \href{http://dx.doi.org/10.1103/PhysRevD.83.125002}{{\em Phys.Rev.}
  {\bfseries D83} (2011) 125002},
  \href{http://arxiv.org/abs/0907.2694}{{\ttfamily arXiv:0907.2694 [hep-th]}}.

\bibitem{Albash:2009wz}
T.~Albash and C.~V. Johnson, ``{Holographic Aspects of Fermi Liquids in a
  Background Magnetic Field},''
  \href{http://dx.doi.org/10.1088/1751-8113/43/34/345405}{{\em J. Phys.}
  {\bfseries A43} (2010) 345405},
\href{http://arxiv.org/abs/0907.5406}{{\ttfamily arXiv:0907.5406 [hep-th]}}.

\bibitem{Basu:2009qz}
P.~Basu, J.~He, A.~Mukherjee, and H.-H. Shieh, ``{Holographic Non-Fermi Liquid
  in a Background Magnetic Field},''
  \href{http://dx.doi.org/10.1103/PhysRevD.82.044036}{{\em Phys. Rev.}
  {\bfseries D82} (2010) 044036},
\href{http://arxiv.org/abs/0908.1436}{{\ttfamily arXiv:0908.1436 [hep-th]}}.

\bibitem{Donos:2011qt}
A.~Donos, J.~P. Gauntlett, and C.~Pantelidou, ``{Spatially modulated
  instabilities of magnetic black branes},''
\href{http://arxiv.org/abs/1109.0471}{{\ttfamily arXiv:1109.0471 [hep-th]}}.

\bibitem{Cacciatori:2009iz}
S.~L. Cacciatori and D.~Klemm, ``{Supersymmetric $AdS_4$ black holes and
  attractors},'' \href{http://dx.doi.org/10.1007/JHEP01(2010)085}{{\em JHEP}
  {\bfseries 01} (2010) 085},
\href{http://arxiv.org/abs/0911.4926}{{\ttfamily arXiv:0911.4926 [hep-th]}}.

\bibitem{Dall'Agata:2010gj}
G.~Dall'Agata and A.~Gnecchi, ``{Flow equations and attractors for black holes
  in N = 2 U(1) gauged supergravity},''
  \href{http://dx.doi.org/10.1007/JHEP03(2011)037}{{\em JHEP} {\bfseries 03}
  (2011) 037},
\href{http://arxiv.org/abs/1012.3756}{{\ttfamily arXiv:1012.3756 [hep-th]}}.

\bibitem{Barisch:2011ui}
S.~Barisch, G.~L. Cardoso, M.~Haack, S.~Nampuri, and N.~A. Obers, ``{Nernst
  branes in gauged supergravity},''
  \href{http://dx.doi.org/10.1007/JHEP11(2011)090}{{\em JHEP} {\bfseries 11}
  (2011) 090},
\href{http://arxiv.org/abs/1108.0296}{{\ttfamily arXiv:1108.0296 [hep-th]}}.

\bibitem{Klebanov:2010tj}
I.~R. Klebanov, S.~S. Pufu, and T.~Tesileanu, ``{Membranes with Topological
  Charge and AdS4/CFT3 Correspondence},''
  \href{http://dx.doi.org/10.1103/PhysRevD.81.125011}{{\em Phys. Rev.}
  {\bfseries D81} (2010) 125011},
\href{http://arxiv.org/abs/1004.0413}{{\ttfamily arXiv:1004.0413 [hep-th]}}.

\bibitem{Denef:2009tp}
F.~Denef and S.~A. Hartnoll, ``{Landscape of superconducting membranes},''
  \href{http://dx.doi.org/10.1103/PhysRevD.79.126008}{{\em Phys. Rev.}
  {\bfseries D79} (2009) 126008},
\href{http://arxiv.org/abs/0901.1160}{{\ttfamily arXiv:0901.1160 [hep-th]}}.

\bibitem{Cvetic:1999xp}
M.~Cvetic {\em et al.}, ``{Embedding AdS black holes in ten and eleven
  dimensions},'' \href{http://dx.doi.org/10.1016/S0550-3213(99)00419-8}{{\em
  Nucl. Phys.} {\bfseries B558} (1999) 96--126},
\href{http://arxiv.org/abs/hep-th/9903214}{{\ttfamily arXiv:hep-th/9903214}}.

\bibitem{Cacciatori:2003kv}
S.~L. Cacciatori, D.~Klemm, and W.~A. Sabra, ``{Supersymmetric domain walls and
  strings in D = 5 gauged supergravity coupled to vector multiplets},'' {\em
  JHEP} {\bfseries 03} (2003) 023,
\href{http://arxiv.org/abs/hep-th/0302218}{{\ttfamily arXiv:hep-th/0302218}}.

\bibitem{Bobev:2010de}
N.~Bobev, A.~Kundu, K.~Pilch, and N.~P. Warner, ``{Supersymmetric Charged
  Clouds in $AdS_5$},'' \href{http://dx.doi.org/10.1007/JHEP03(2011)070}{{\em
  JHEP} {\bfseries 03} (2011) 070},
\href{http://arxiv.org/abs/1005.3552}{{\ttfamily arXiv:1005.3552 [hep-th]}}.

\bibitem{Cvetic:2000nc}
M.~Cvetic, H.~Lu, C.~Pope, A.~Sadrzadeh, and T.~A. Tran, ``{Consistent SO(6)
  reduction of type IIB supergravity on S**5},''
  \href{http://dx.doi.org/10.1016/S0550-3213(00)00372-2}{{\em Nucl.Phys.}
  {\bfseries B586} (2000) 275--286},
  \href{http://arxiv.org/abs/hep-th/0003103}{{\ttfamily arXiv:hep-th/0003103
  [hep-th]}}.

\bibitem{Chong:2004ce}
Z.~W. Chong, H.~Lu, and C.~N. Pope, ``{BPS geometries and AdS bubbles},''
  \href{http://dx.doi.org/10.1016/j.physletb.2005.03.050}{{\em Phys. Lett.}
  {\bfseries B614} (2005) 96--103},
\href{http://arxiv.org/abs/hep-th/0412221}{{\ttfamily arXiv:hep-th/0412221}}.

\bibitem{Liu:2007rv}
J.~T. Liu, H.~Lu, C.~N. Pope, and J.~F. Vazquez-Poritz, ``{New supersymmetric
  solutions of N=2, D=5 gauged supergravity with hyperscalars},''
  \href{http://dx.doi.org/10.1088/1126-6708/2007/10/093}{{\em JHEP} {\bfseries
  10} (2007) 093},
\href{http://arxiv.org/abs/0705.2234}{{\ttfamily arXiv:0705.2234 [hep-th]}}.

\bibitem{Duff:1999gh}
M.~J. Duff and J.~T. Liu, ``{Anti-de Sitter black holes in gauged N = 8
  supergravity},'' \href{http://dx.doi.org/10.1016/S0550-3213(99)00299-0}{{\em
  Nucl. Phys.} {\bfseries B554} (1999) 237--253},
\href{http://arxiv.org/abs/hep-th/9901149}{{\ttfamily arXiv:hep-th/9901149}}.

\bibitem{Warner:1983vz}
N.~P. Warner, ``{Some new extrema of the scalar potential of gauged N=8
  supergravity},''
\href{http://dx.doi.org/10.1016/0370-2693(83)90383-0}{{\em Phys. Lett.}
  {\bfseries B128} (1983) 169}.

\bibitem{Bobev:2010ib}
N.~Bobev, N.~Halmagyi, K.~Pilch, and N.~P. Warner, ``{Supergravity
  Instabilities of Non-Supersymmetric Quantum Critical Points},''
  \href{http://dx.doi.org/10.1088/0264-9381/27/23/235013}{{\em Class. Quant.
  Grav.} {\bfseries 27} (2010) 235013},
\href{http://arxiv.org/abs/1006.2546}{{\ttfamily arXiv:1006.2546}}.

\bibitem{Gauntlett:2006ux}
J.~P. Gauntlett, O.~A.~P. Mac~Conamhna, T.~Mateos, and D.~Waldram, ``{AdS
  spacetimes from wrapped M5 branes},''
  \href{http://dx.doi.org/10.1088/1126-6708/2006/11/053}{{\em JHEP} {\bfseries
  11} (2006) 053},
\href{http://arxiv.org/abs/hep-th/0605146}{{\ttfamily arXiv:hep-th/0605146}}.

\bibitem{Kim:2006qu}
N.~Kim and J.-D. Park, ``{Comments on AdS(2) solutions of D = 11
  supergravity},'' \href{http://dx.doi.org/10.1088/1126-6708/2006/09/041}{{\em
  JHEP} {\bfseries 09} (2006) 041},
\href{http://arxiv.org/abs/hep-th/0607093}{{\ttfamily arXiv:hep-th/0607093}}.

\end{thebibliography}\endgroup

\end{document}